\documentclass[11pt]{article}

\usepackage[a4paper,margin=1in]{geometry}
\usepackage[T1]{fontenc}
\usepackage{lmodern}
\usepackage{microtype}
\usepackage{setspace}
\usepackage{graphicx}
\graphicspath{{figures/}}
\usepackage{float}
\usepackage[font=small,labelfont=bf,labelsep=period,justification=centering]{caption}

\usepackage{booktabs}
\usepackage{array}
\usepackage{tabularx}
\usepackage{longtable}
\newcolumntype{L}[1]{>{\raggedright\arraybackslash}p{#1}}

\usepackage{enumitem}
\setlist{itemsep=2pt,topsep=4pt}
\usepackage{amsmath,amssymb}

\usepackage[explicit]{titlesec}
\titleformat{\section}{\Large\bfseries}{\thesection}{0.6em}{#1}
\titleformat{\subsection}{\large\bfseries}{\thesubsection}{0.5em}{#1}
\titleformat{\subsubsection}{\normalsize\bfseries}{\thesubsubsection}{0.4em}{#1}
\titlespacing*{\section}{0pt}{1.4\baselineskip}{0.6\baselineskip}
\titlespacing*{\subsection}{0pt}{1.0\baselineskip}{0.4\baselineskip}
\titlespacing*{\subsubsection}{0pt}{0.8\baselineskip}{0.3\baselineskip}

\usepackage[hidelinks,breaklinks=true,colorlinks=false]{hyperref}
\usepackage{url}
\renewenvironment{abstract}
  {\begin{center}\bfseries\large Abstract\end{center}%
   \begin{quotation}\small\noindent\ignorespaces}
  {\end{quotation}}

\title{\bfseries C8s: A Confidential Kubernetes Architecture}
\author{%
  \begin{tabular}{c@{\hskip 2em}c@{\hskip 2em}c}
    Amean Asad\thanks{Equal contribution.} & Patrick McClurg\footnotemark[1] & Jo\~ao Andrade \\
    \texttt{amean@confidential.ai} & \texttt{patrick@confidential.ai} & \texttt{joao@confidential.ai}
  \end{tabular}
  \\[1.4em]
  \href{https://confidential.ai}{Confidential.ai}
}
\date{April 2026}

\begin{document}
\maketitle

\begin{abstract}
This paper presents C8s, a confidential computing architecture for Kubernetes that provides cryptographically rooted confidentiality, integrity, and verifiability guarantees for Kubernetes clusters from infrastructure operators. These guarantees are cryptographically provable to any independent third party verifier. The architecture is built on hardware Trusted Execution Environments (TEEs), specifically AMD SEV-SNP, Intel TDX, and NVIDIA Confidential Computing support, to establish an attestation-rooted trust boundary around confidential VMs. This design is compatible with managed Kubernetes services such as Amazon EKS, Google GKE, and Microsoft AKS, where the control plane cannot be attested. Under this boundary, three groups gain guarantees that are absent from conventional deployments. Data and artifact owners can deploy sensitive workloads and proprietary artifacts on third-party infrastructure without risking exfiltration. Compute providers can offer execution services without revealing workloads to cloud operators. End users can submit requests that remain opaque to all parties except the attested TEE processing them. Representative workloads include AI inference, securing AI model weights, and training or fine-tuning on sensitive data.
\end{abstract}

\section{Introduction}

\subsection{Motivation}

Running sensitive workloads on third-party infrastructure presents a three-sided confidentiality problem. Artifact owners invest substantial resources in proprietary assets such as trained model weights, curated datasets, cryptographic keys, or proprietary algorithms, and risk exfiltration when those assets are deployed on infrastructure they do not control. Compute providers running execution infrastructure need assurance that cloud providers cannot observe their workloads. End users submitting sensitive requests, in healthcare, legal, financial, and national security contexts, require guarantees that neither the compute provider nor the infrastructure operator can inspect their inputs or outputs.

Today, the dominant mitigation for all three concerns is contractual. Model providers sign licensing agreements restricting weight access. Cloud providers publish compliance certifications. Service providers implement access controls. None of these mechanisms are cryptographically rooted. A sufficiently privileged insider, such as a cloud engineer with hypervisor access, a Kubernetes administrator with credentials, or a compromised node with root access, can read data from memory, intercept network traffic, or exfiltrate model weights from storage.

Encryption at rest and encryption in transit address portions of this problem but leave a fundamental gap during computation where data is plaintext in memory. Workloads, user data, and proprietary intellectual property all reside in memory cleartext, readable by any party with sufficient host access.

Kubernetes is a dominant orchestration layer for workloads in production, and therefore a generalized confidential architecture at the cluster level targets a substantial fraction of existing deployed compute infrastructure. Rather than requiring workloads to be rewritten against a specialized confidential runtime, the approach here preserves the Kubernetes surface area that operators and developers already know, and any workload already running on a Kubernetes cluster can inherit these guarantees with minimal modification.

\begin{figure}[H]
\centering
\includegraphics[width=0.92\linewidth]{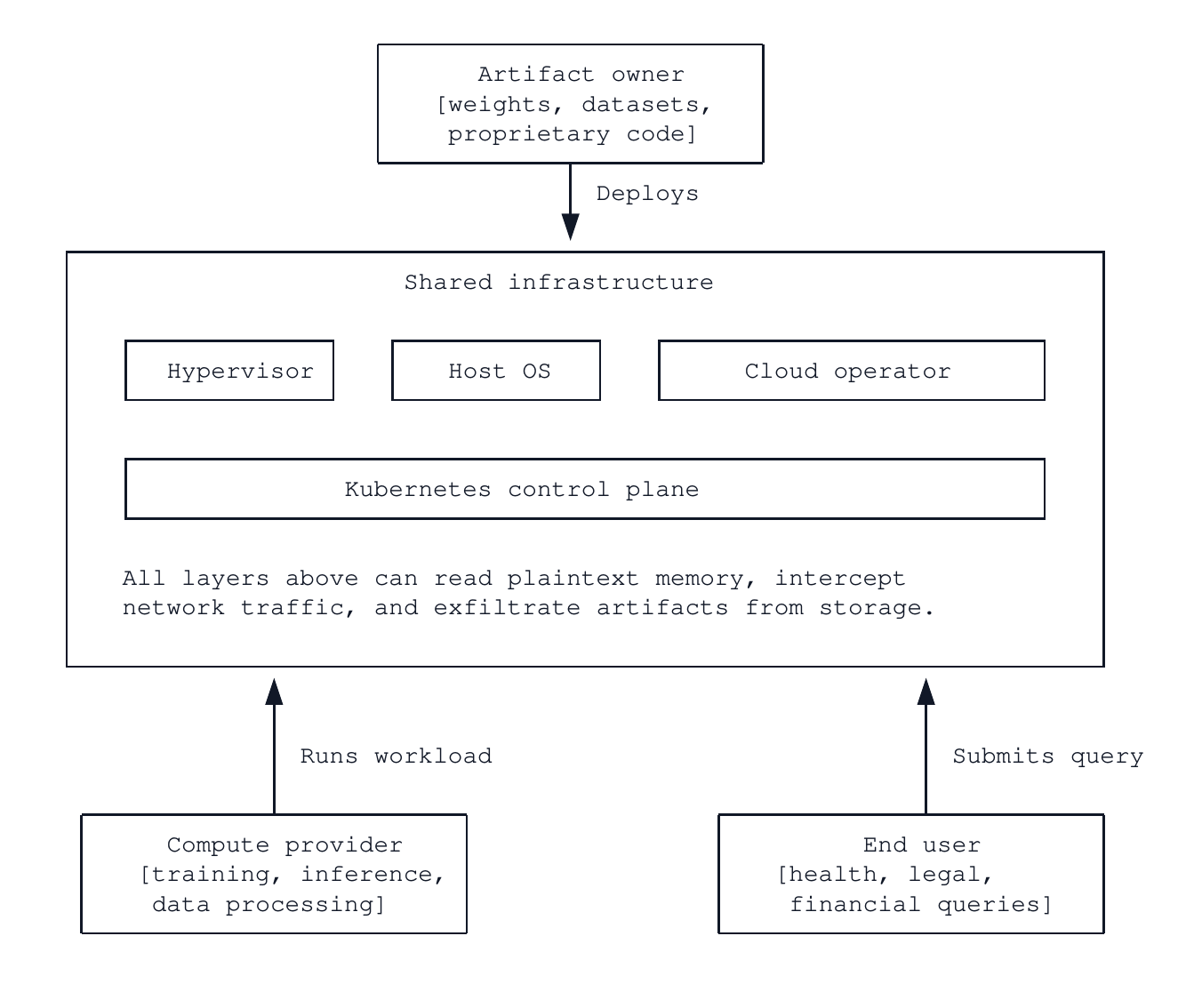}
\caption{The three-sided trust problem. Artifact owners, compute providers, and end users share infrastructure none of them controls; every confidentiality boundary between them is enforced by contract rather than by cryptography.}
\label{fig:three-sided}
\end{figure}

\subsection{Problem Statement}

The core challenge is to construct a deployment architecture for sensitive workloads on Kubernetes in which:

\begin{enumerate}
\item Workload inputs, outputs, and intermediate computation state are cryptographically protected from infrastructure operators.
\item Sensitive artifacts, such as model weights, datasets, cryptographic keys, or proprietary code, are encrypted at rest and decrypted only within hardware-attested environments.
\item Clients and customers can independently verify, through cryptographic attestation, that their data and workloads are processed exclusively within trusted hardware running expected code.
\item The architecture operates on managed Kubernetes services where the control plane is outside the operator's control and cannot be attested.
\end{enumerate}

\subsection{Scope}

This document describes the architecture, threat model, and core components of the C8s (Confidential Kubernetes) platform developed by Confidential. It covers the trust model, the attestation and certificate issuance flow, the client connection protocol, and the design decisions that enable compatibility with untrusted control planes across a general class of sensitive workloads. The attested build system (Kettle) is referenced but not detailed; it will be the subject of a separate publication.

\section{Background}

\subsection{Trusted Execution Environments}

A Trusted Execution Environment (TEE) is a hardware-enforced isolation boundary that provides four security properties:

\begin{itemize}
\item \textbf{Confidentiality.} Data within the TEE is encrypted at the hardware level. Memory pages are encrypted with keys managed by the CPU's secure processor. The hypervisor, host operating system, and DMA-capable devices cannot decrypt TEE memory contents.
\item \textbf{Integrity.} Replay attacks, data corruption, and memory remapping are protected by hardware. This maintains the integrity of the initialized software in the TEE.
\item \textbf{Verifiability.} The TEE produces a cryptographic measurement (a hash digest) of the software loaded at launch. This measurement covers firmware, kernel, initial filesystem state, and user space, and is computed by the CPU's secure processor before the guest operating system begins execution.
\item \textbf{Attestation.} The TEE generates signed attestation reports (termed \emph{Evidence} in the RATS architecture~\cite{ref:rats}) binding the launch measurement to a hardware-rooted signing key. These reports are signed by keys that chain to the hardware manufacturer's root of trust and cannot be forged by any software on the host. The hardware manufacturer takes the role of the \emph{Endorser} in RATS terminology, vouching for the platform via the chain of certificates rooted at its public key.
\end{itemize}

\begin{figure}[H]
\centering
\includegraphics[width=0.98\linewidth]{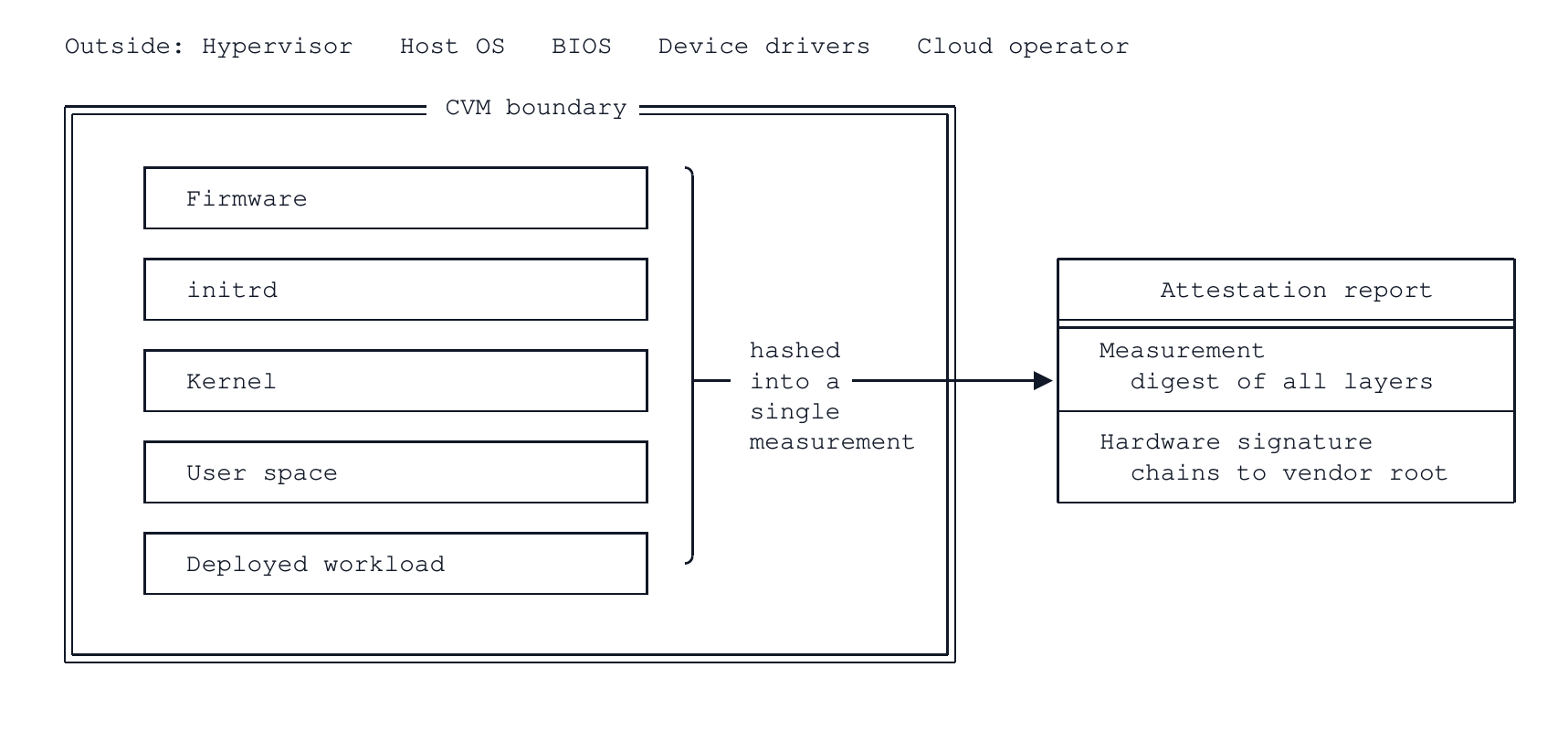}
\caption{The CVM boundary and attestation report. A digest covering firmware, initrd, kernel, user space, and the deployed workload is signed by hardware-rooted keys and produced as the attestation report; nothing outside the boundary can observe or modify the measured layers.}
\label{fig:cvm-boundary}
\end{figure}

AMD Secure Encrypted Virtualization with Secure Nested Paging (SEV-SNP)~\cite{ref:amd-sev-snp} and Intel Trust Domain Extensions (TDX)~\cite{ref:intel-tdx} implement these properties at the virtual machine level. In both architectures, the entire VM operates as a Confidential VM (CVM), and all guest memory is encrypted with keys the hypervisor never possesses.

NVIDIA Confidential Computing (CC) mode~\cite{ref:nvidia-cc} extends hardware-level memory encryption to GPU VRAM. When a GPU operates in CC mode, data crossing the PCIe bus between the CPU TEE and GPU is encrypted and integrity-protected via bounce buffers, and the GPU isolates its protected memory and state from the host hypervisor and OS. The GPU is able to generate its own attestation confirming that it is operating in CC mode. To use CC mode on GPUs, the GPU must be paired with a CC-capable CPU. In that setup, the CPU is the attestation trust anchor.

For GPU-accelerated workloads, GPU CC mode is significant because the working set, whether that consists of model parameters, training data, intermediate activations, or other sensitive state, resides in VRAM for the lifetime of the compute process. Without GPU CC mode, VRAM is the one location where plaintext data is not hardware-protected. NVIDIA H100 and later GPUs with CC mode enabled close this gap.

The C8s architecture supports AMD SEV-SNP and Intel TDX for CPU-level CVM attestation, and NVIDIA GPUs operating in CC mode for GPU-level attestation. The same attestation machinery, including CDS verification, NRI policy enforcement, raTLS mesh, and client-side encryption, applies uniformly across CPU-only and GPU-accelerated nodes. On GPU-equipped nodes, the GPU's attestation evidence is included in the node's combined attestation report, and the CDS verifies it alongside the CPU measurement.

\subsection{Standard Kubernetes Architecture}

A Kubernetes deployment is organized as a cluster of machines, called nodes, that collectively run containerized workloads. Each node is a physical or virtual host contributing CPU, memory, and optionally GPU resources to the cluster. Workloads run inside pods, which are the smallest deployable unit in Kubernetes. Each pod is a group of one or more containers scheduled together on the same node and sharing its network and storage namespaces. A control plane, running separately from the worker nodes, accepts desired-state declarations from operators and schedules pods onto nodes that can satisfy their resource requirements.

A typical production deployment on this substrate comprises one or more application pods fronted by a routing layer. External traffic arrives at a load balancer or ingress controller, which terminates TLS and forwards requests into the cluster. The routing layer, whether a dedicated router, a service mesh gateway, or a Kubernetes Service, distributes requests across available pods via Kubernetes service discovery. Application pods run the workload with dedicated CPU, GPU, or memory resources, pulling container images from a registry and loading any required data or artifacts from persistent volumes, object storage, or configuration at startup.

\begin{figure}[H]
\centering
\includegraphics[width=0.78\linewidth]{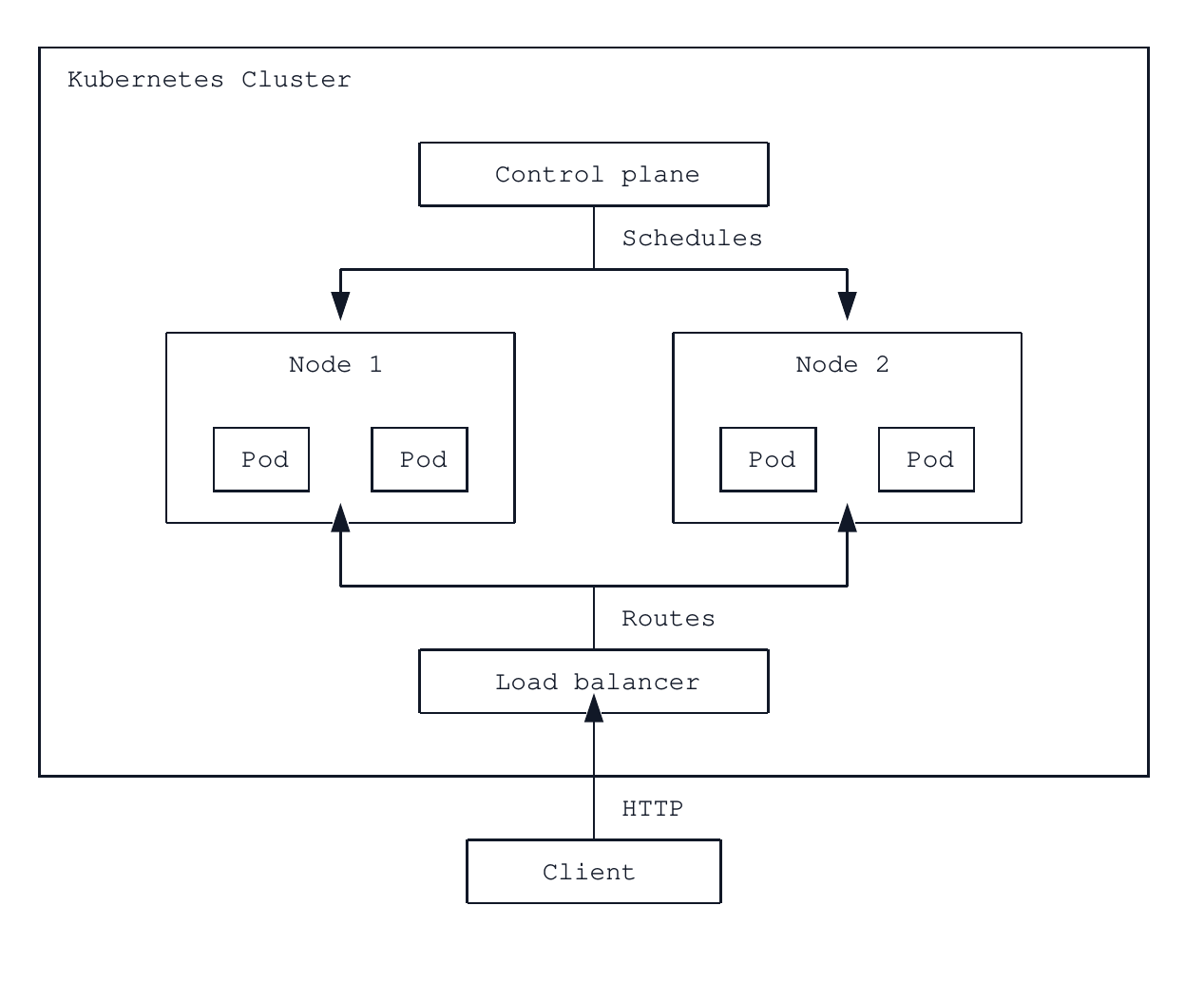}
\caption{Standard Kubernetes topology. The control plane schedules pods onto nodes; an external load balancer distributes client traffic. No confidentiality enforcement is present.}
\label{fig:std-k8s}
\end{figure}

\subsection{Related Work}

Prior work on confidential computing in Kubernetes, including Kata Containers~\cite{ref:kata} and the Confidential Containers (CoCo) project~\cite{ref:coco}, has focused on per-pod isolation by running each pod in its own micro-VM with an independent attestation. These projects establish much of the ecosystem's shared machinery, including confidential container runtimes, per-pod attestation agents, and key broker services that release secrets to individually attested containers. Single-tenant large-scale deployments, where container-level isolation is not strictly required, have historically been a poor fit for this model because every workload has to be adapted to run in a confidential container runtime, and the integration cost falls on every Kubernetes deployment that does not already run its containers inside VMs. Edgeless Systems' Constellation~\cite{ref:constellation} takes the approach of wrapping an entire Kubernetes cluster, control plane and workers, in confidential VMs with a cluster-wide attestation and integrated key management.

The approach C8s takes is to have multiple configuration options that widen the deployment options and lower integration costs. The C8s architecture treats the choice of isolation boundary as a deployment-time configuration. Available options range from node level attestation, to per pod attestation, or hybrid configurations within a cluster. When the boundary is drawn at the pod, per-pod confidential VM lifecycle is orchestrated by the \textbf{C8s pod runtime}, which presents a direct per-pod attestation to the same CDS. A user or customer can verify these deployment configurations independently through the trust boundary that C8s draws. This deployment flexibility makes C8s compatible with existing managed Kubernetes services such as GKE, AKS, and EKS.

\subsection{Multi-Recipient Encryption}
\label{sec:mre}

For deployments that opt into client-side encryption (\S\ref{sec:client-mre}), the architecture needs a mechanism for encrypting payloads to a pool of TEEs such that any single TEE in the pool can decrypt the payload. This serves two purposes. The first is availability, since TEE instances may be terminated or rescheduled at any time without requiring client-side key refresh. The second is operational simplicity, since clients encrypt once per request rather than per-TEE. The natural fit is a hybrid scheme that encrypts the body once under a fresh symmetric key and then wraps that key to each recipient's public key, so the body cost is independent of pool size and only the header grows.

Several schemes satisfy this requirement. Elliptic Curve Integrated Encryption Scheme (ECIES)~\cite{ref:ecies} supports multi-recipient wrapping by encrypting a symmetric data key to each recipient's public key. Broadcast encryption schemes~\cite{ref:broadcast} offer similar semantics with different tradeoff profiles. The current C8s reference implementation uses AGE (Actually Good Encryption)~\cite{ref:age}, a file encryption format designed for simplicity and composability; the protocol is not bound to AGE and can be substituted (e.g., for threshold, hybrid, or post-quantum schemes) without changing the rest of the architecture.

AGE supports multiple recipients via a stanza-based header format. A file is encrypted with a single symmetric key, and each recipient receives an independently decryptable copy of that key wrapped to their public key. The encrypted payload body is identical regardless of the number of recipients; only the header grows, by approximately 64 bytes per additional recipient. For a pool of 100 TEEs, the header overhead is approximately 6.4 KB, negligible relative to payloads. Two practical properties motivated the choice. The header format supports custom recipient types via a plugin system, allowing routing metadata (such as hostnames) to ride alongside cryptographic material, and the plugin architecture itself provides crypto-agility for the schemes mentioned above. The routing hints embedded in recipient stanzas are discussed further in \S\ref{sec:encrypted-ingress}.

The CDS (introduced in detail in \S\ref{sec:cds}) signs the manifest of TEE public keys distributed to clients, so clients can verify the authenticity of the key set using standard certificate chain validation without performing individual attestation verification on every request.

\section{Threat Model}

\subsection{Assets Under Protection}

The architecture protects three categories of assets:

\begin{enumerate}
\item \textbf{Workload inputs and outputs.} The data submitted to the workload and the results it returns. Examples include user queries and generated tokens in LLM inference, patient records and diagnostic outputs in healthcare analytics, and transaction data and risk scores in financial pipelines.
\item \textbf{Sensitive artifacts.} Proprietary assets loaded into the workload at runtime, such as trained model weights, curated datasets, cryptographic keys, or proprietary code.
\item \textbf{Intermediate computation state.} Caches, activations, and working memory generated during execution. Examples include KV cache entries and attention activations in LLM inference, intermediate aggregates in analytics workloads, and partial witnesses in cryptographic proving.
\end{enumerate}

\subsection{Trust Model}
\label{sec:trust-model}

The trust boundary encompasses the hardware manufacturer, the code running inside attested TEEs, and the CDS (which is itself attested and explicitly trusted as the root of the certificate chain). All other parties are untrusted.

\begin{table}[H]
\centering
\small
\begin{tabularx}{\linewidth}{X X}
\toprule
\textbf{Trusted} & \textbf{Untrusted} \\
\midrule
Hardware manufacturer (AMD, Intel) & Cloud provider / hypervisor \\
CPU/GPU hardware and firmware & Host OS, BIOS, device drivers \\
Physical hardware host (datacenter, colo) & Network infrastructure \\
Code inside TEE (measured at launch) & Kubernetes control plane \\
CDS (attested, trusted root of certificate chain) & Operators with privileged access \\
& Storage backends \\
\bottomrule
\end{tabularx}
\caption{Components inside vs.\ outside the trust boundary.}
\end{table}

The trust model places the workload provider's own operators on the untrusted side. This is deliberate, because the architecture offers guarantees that do not depend on the operational security of the party running the infrastructure.

One physical-security assumption is required and worth stating explicitly. The physical hardware host (the datacenter, colo, or party that racks and cables the machine) is trusted not to mount physical attacks against the silicon, including memory-bus probing, DIMM substitution, and JTAG or debug-port access. In cloud deployments the physical hardware host and the hypervisor operator are usually the same company (e.g., AWS, Azure, GCP). The model trusts that company in its physical hardware host role while explicitly distrusting it in its hypervisor-operator role. Recent results such as TEE.Fail (2025), Battering RAM (2025), and BadRAM (2024) demonstrate that current TEE designs are vulnerable to adversaries with physical access to running hardware. Protection against such adversaries is out of scope and is treated further in \S\ref{sec:attacks-not}.

\subsection{Attacks Addressed}

\begin{itemize}
\item \textbf{Memory snooping.} The hypervisor, host OS, and DMA-capable devices cannot read CVM memory. Data in DRAM, queries in process memory, and cache entries are encrypted with keys the hypervisor never possesses.
\item \textbf{Memory tampering.} Replay attacks, corruption, and memory remapping are detected by the hardware. The guest always observes the data it last wrote.
\item \textbf{Memory snapshots.} The TEE protects against memory snapshots taken by the hypervisor while the VM is running. The hypervisor can capture an image of the VM's memory pages, but the contents are encrypted with keys it does not possess, rendering the snapshot useless for data extraction.
\item \textbf{Code substitution.} Attestation binds a cryptographic measurement to the software running in the TEE. Substituting a malicious binary changes the measurement, which is detected during attestation verification.
\item \textbf{Man-in-the-middle (internal).} All intra-cluster traffic uses mutual TLS (mTLS) with certificates issued by the Verifier. Only workloads that have passed attestation possess valid certificates for mTLS.
\item \textbf{Man-in-the-middle (external).} Clients verify the ingress's attestation against a signed manifest and the freshness beacon (\S\ref{sec:default-tls}) before trusting the TLS session. The TLS certificate's public key is bound by the attestation report to a key generated inside the ingress's TEE, so a forged certificate cannot be presented with valid attestation.
\item \textbf{Unauthorized workload injection.} Workload gating happens inside the trust boundary: in pod-level deployments, the in-pod measured policy or per-customer signing key (\S\ref{sec:image-gap}) rejects unauthorized images before they execute; in node-level deployments, the NRI image policy enforcer (\S\ref{sec:nri}) rejects container images whose digests do not appear in a signed allow-list. Both paths are independent of control plane directives.
\item \textbf{Key exfiltration.} Decryption keys for sensitive artifacts are released by the CDS only to workloads that pass attestation, and exist exclusively in the pod's hardware-encrypted memory. They are never stored in Kubernetes Secrets, environment variables, or any location accessible to the control plane.
\end{itemize}

\subsection{Attacks Not Addressed}
\label{sec:attacks-not}

\begin{itemize}
\item \textbf{Side-channel attacks.} Microarchitectural timing attacks (cache timing, branch prediction, page-level access patterns) are not fully mitigated by current CVM implementations. While SEV-SNP and TDX address integrity-based attacks demonstrated against earlier SEV generations~\cite{ref:severed}, cache-based and ciphertext side channels remain an active area of research.
\item \textbf{Denial of service.} A malicious operator or compromised control plane can refuse to schedule workloads, terminate VMs, or disrupt network connectivity. TEEs guarantee confidentiality and integrity, not availability.
\item \textbf{Application-layer vulnerabilities.} A bug in the workload code is exploitable inside the TEE. The hardware protects the execution environment, not the application logic.
\item \textbf{Application-layer extraction.} Repeated querying to distill a model, reconstruct a dataset, or infer proprietary logic from workload responses is an application-layer concern outside the scope of hardware-level protections.
\item \textbf{Physical attacks.} Active probing of the memory bus, DIMM substitution, debug-port access, and similar attacks requiring physical access to running hardware are out of scope. The physical hardware host is trusted under \S\ref{sec:trust-model}; an adversary with sustained physical access to the machine is outside the threat model. Recent attacks in this category include TEE.Fail (2025)~\cite{ref:teefail}, Battering RAM (2025)~\cite{ref:battering-ram}, and BadRAM (2024)~\cite{ref:badram}.
\end{itemize}

\section{Architecture}

\subsection{Trust Boundary}

The central construct of the architecture is a \textbf{trust boundary}, a cryptographically enforced perimeter separating components that the architecture treats as trusted from those that it does not. Inside the boundary, components run in hardware-attested environments and hold identities issued on the basis of that attestation. Outside the boundary, components are treated as potentially adversarial, regardless of who operates them, and no confidentiality or integrity guarantee depends on their correct behavior.

The trust boundary is defined entirely by attestation. A component is inside the boundary if and only if three conditions hold:

\begin{enumerate}
\item Its code or image is measurable, meaning a cryptographic digest can be computed over the software that will execute.
\item Its execution environment supports confidential computing, either as a CVM itself, or as a container running inside an already-attested CVM.
\item A verification chain exists from the hardware root of trust to the component's runtime identity such that a remote Verifier can confirm the environment is genuine and the running code matches an expected measurement.
\end{enumerate}

Any component that fails one of these conditions sits outside the boundary. Network infrastructure, storage backends, and managed Kubernetes control planes all fail at least the second condition, since the operator does not run the code in a TEE and cannot attest it. These components remain part of the deployment but are stripped of any role in confidentiality enforcement. The architecture depends on them for liveness and orchestration, not for secrecy or integrity.

For components that cannot be placed inside the boundary, two strategies apply:

\begin{enumerate}
\item \textbf{Do not depend on them for confidentiality.} The control plane schedules workloads, but an image policy enforcer on each node independently verifies every container image digest against a signed allow-list. The control plane can request arbitrary workloads; the node rejects those not authorized.
\item \textbf{Encrypt before exposure.} Data passing through untrusted components is ciphertext. Storage backends hold encrypted data. The pod network carries only mTLS-encrypted traffic. Client traffic is anchored to an attested ingress over TLS (and may additionally be multi-recipient-encrypted for clients that opt into end-to-end-to-pod confidentiality).
\end{enumerate}

\subsection{Design Principles}
\label{sec:design-formula}

With the trust boundary defined, the architecture applies a repeatable formula to every component placed inside it. The formula operationalizes the boundary by making each component independently measurable, identifiable, and verifiable, so that trust composes cleanly across the system.

For every component that should go inside the trust boundary, we secure it by implementing the following steps:

\begin{enumerate}
\item \textbf{Encrypt the runtime.} The component runs in hardware-encrypted memory.
\item \textbf{Measure the code.} A cryptographic digest of the running software is computed by hardware at launch.
\item \textbf{Bind identity to measurement.} Credentials are issued only after the measurement is verified against a known-good value.
\item \textbf{Verify before connecting.} Peers require attestation-rooted identity before accepting connections.
\item \textbf{Secure the egress.} All traffic leaving the component is encrypted and authenticated to verified destinations.
\end{enumerate}

Steps 1 to 4 enable external parties to verify they are communicating with a genuine TEE running expected code. Step 5 ensures the component does not leak data to unverified infrastructure. The component sections that follow describe each piece of the system once, but the discipline they share is this five-step formula applied at a different layer of the stack; a reader asking how a given component becomes confidential can check it against these five steps and find the answer.

\subsection{Component Overview}

The C8s platform introduces the following components:

\begin{itemize}
\item \textbf{CVM node image.} A VM image that boots as a confidential VM on supported TEE hardware. The hardware's secure processor measures the initial state at launch. When GPU CC mode is available, the measurement additionally attests the GPU, producing a combined attestation report covering the full compute path. All processes running inside inherit hardware memory encryption.
\item \textbf{Certificate Distribution Service (CDS).} The root of the certificate chain within the boundary. The CDS verifies attestation reports from CVMs, checks measurements against an allow-list, and issues per-pod raTLS mesh certificates to attested workloads. It also issues freshness beacons on request, letting external clients verify that an attested ingress's recent attestation is current (\S\ref{sec:default-tls}). External-facing TLS certificates at the ingress are obtained through ordinary public-CA mechanisms bound to a TEE-internal key (out of scope here), so the CDS itself does not issue any externally facing certificate. The CDS runs inside a CVM and is itself attestable, and acts as the attestation-gated broker for application-layer secrets including model decryption keys (\S\ref{sec:key-broker}).
\item \textbf{Attestation service.} An in-CVM component that generates attestation reports and presents them to a verifying party. It runs inside every attested CVM in the cluster, as a DaemonSet inside the node's CVM in node-level deployments and as a per-pod component inside the pod's CVM in pod-level deployments. The service requests the hardware-signed report from the CVM, binds it to the workload's ephemeral public key, and submits it to the CDS for appraisal; in node-level deployments it additionally composes per-pod workload digests on top of the node's CVM report so each pod presents a distinct report.
\item \textbf{C8s pod runtime.} A pod-level runtime layer that launches each pod as its own confidential VM. The runtime produces a pre-computable launch digest covering the pod's boot chain and vCPU initial state, so every pod attests directly to the CDS at start with no dependence on node-level evidence. Workload gating moves into the pod's measured configuration because the host does not see the post-boot container pull. Used wherever the trust boundary is drawn at the pod rather than the node.
\item \textbf{NRI image policy enforcer.} A Node Resource Interface~\cite{ref:nri} plugin that intercepts every container launch request on each node. It checks the image digest against a signed policy manifest. If the digest is not in the allow-list, the container does not start. This is node-level enforcement, independent of control plane directives. It applies where the trust boundary is drawn at the node; where the boundary is drawn at the pod, the equivalent gating is enforced inside the pod's measured configuration.
\item \textbf{Kettle (attested build system).} A build pipeline that executes inside a TEE, producing container images with deterministic, reproducible digests. The build environment is attested, binding the inputs (source code, dependencies) to the output (image digest) with a hardware-signed attestation report. Kettle is out of scope for this paper and will be detailed in a separate publication.
\item \textbf{raTLS mesh.} A mutual TLS overlay that encrypts all traffic crossing the trust boundary. Every pod holds its own CDS-issued certificate, issued only after the pod's attestation evidence verifies against policy: a direct per-pod attestation where the boundary sits at the pod (default), or composed node-and-workload evidence where the boundary sits at the node. The mesh terminates at the pod, not the node: the pod network, which is untrusted, sees only ciphertext between attested pod endpoints.
\item \textbf{Attestation-aware client.} A client library that fetches a CDS-signed manifest of attested identities (the ingress and the destination TEE pool), verifies attestation evidence either independently or via the CDS signature, and uses the verified material to anchor a TLS connection to the ingress (default) or to encrypt payloads with multi-recipient encryption (optional, \S\ref{sec:client-mre}).
\item \textbf{Ingress router.} A load balancer running an attested image inside the trust boundary. By default it terminates a public-CA TLS connection bound to its TEE-internal key (clients verify the attached attestation against the CDS-signed manifest and the freshness beacon) and forwards over raTLS to the destination pod. An Encrypted Ingress Router variant (\S\ref{sec:encrypted-ingress}) handles client-side multi-recipient encrypted payloads by routing on header metadata without decrypting the body.
\item \textbf{Secrets Manager Proxy.} A slim service inside the trust boundary that brokers access to enterprise secrets managers sitting outside the trust boundary, releasing secrets only on presentation of a valid CDS-issued mesh identity. Role-based access can be gated on measurements from the attestation. See \S\ref{sec:key-broker} for the integration pattern.
\end{itemize}

\subsection{Architecture Diagram}

The following pair of diagrams illustrates what changes when a standard Kubernetes cluster becomes confidential. The top panel is a conventional deployment, and the bottom panel is the same workload under C8s. The trust boundary is hardware-enforced and shown with a double wall, while the control plane stays outside it.

\begin{figure}[H]
\centering
\includegraphics[width=0.85\linewidth]{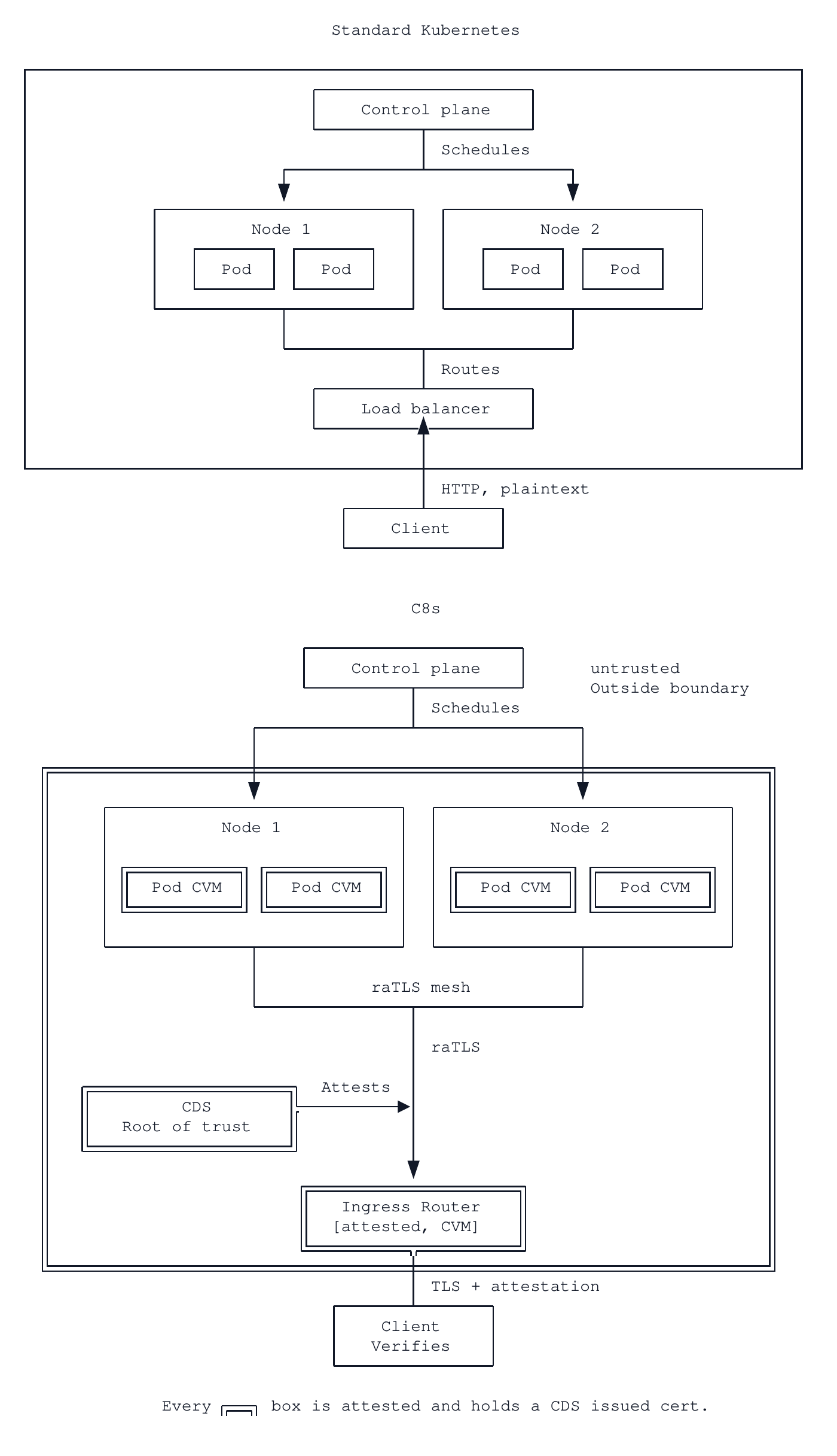}
\caption{Standard Kubernetes vs.\ C8s. The control plane remains outside the trust boundary in both cases; under C8s, every node and the ingress are CVMs holding CDS-issued identities, and intra-cluster traffic runs over the raTLS mesh.}
\label{fig:before-after}
\end{figure}

\section{Component Design}

The pod is the unit of confidentiality, identity, and attestation in C8s. Every component described in this section is in service of that arrangement, and the CVM boundary around the pod is a deployment-time choice rather than a fixed feature of the architecture. The sections below introduce each component once, in the order a reader needs them to understand how a request reaches an attested pod and leaves again.

\subsection{Pod-level CVM (default)}
\label{sec:pod-cvm}

C8s draws the trust boundary at the pod by default. Each pod boots inside its own confidential VM and carries its own launch measurement. Per-pod lifecycle in this configuration is orchestrated by the \textbf{C8s pod runtime}, which builds on Kata Containers~\cite{ref:kata} machinery for the per-pod VM lifecycle and extends it with a pre-computable measured boot, direct CDS integration, and the signed-workload discipline described below. The worker node itself is no longer the attested unit. It hosts the runtime and proxies pod lifecycle to per-pod CVMs, but carries no workload trust of its own. A node-level boundary is also supported (\S\ref{sec:node-cvm}) for deployments where it is preferable; \S\ref{sec:choosing-boundary} discusses when to choose each.

The same design formula from \S\ref{sec:design-formula} applies per-pod. Each pod runs in hardware-encrypted memory. The pod runtime produces a launch digest that covers the pod's entire boot chain and vCPU initial state, so the digest can be recomputed ahead of time and checked against an allow-list without inspecting the running system. The CDS verifies the per-pod attestation report directly and issues a certificate bound to the pod's ephemeral key. Peers verify that certificate before accepting connections, and egress is encrypted via the raTLS mesh exactly as it is at the node level. The surrounding architecture (CDS, raTLS, attestation-gated key release, client-side encryption) does not change; only the evidence shape does.

\begin{figure}[H]
\centering
\includegraphics[width=0.92\linewidth]{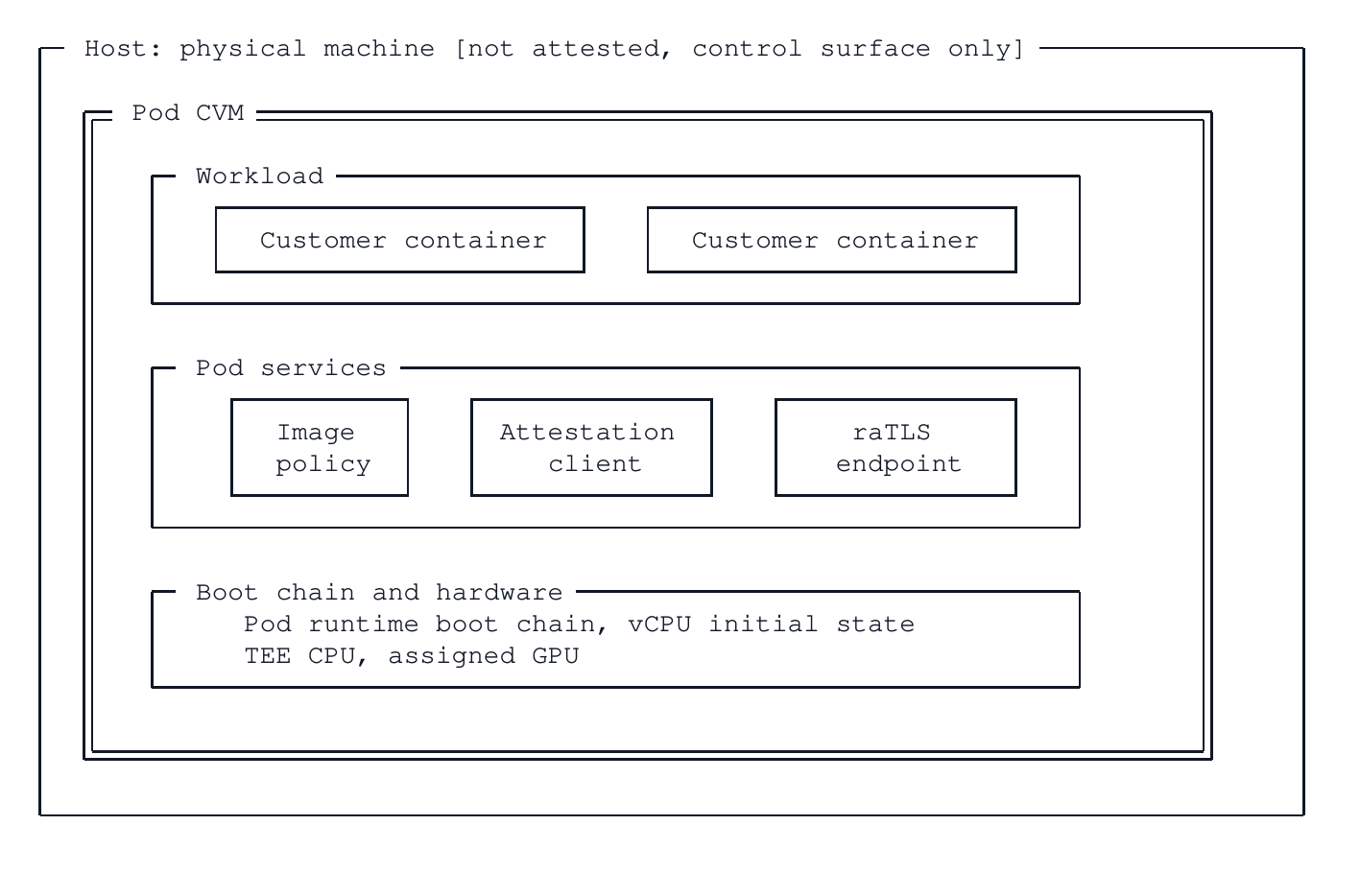}
\caption{Pod-level CVM anatomy. The host is an ordinary VM acting as a control surface; every pod boots inside its own confidential VM whose measured boot chain covers the in-pod services that gate image pulls and present mesh identity.}
\label{fig:pod-cvm}
\end{figure}

\subsubsection{Image-gap Mechanisms}
\label{sec:image-gap}

A per-pod launch digest covers the software that measures at boot, but not the container image, which is pulled from a registry after the pod is already running. This is the same gap that NRI enforcement closes at the node level; when the boundary moves to the pod, the host can no longer observe the pull, so the gating has to move inside. C8s supports three mechanisms, chosen per deployment or combined:

\begin{enumerate}
\item \textbf{Measured image pinning.} A policy baked into the pod's measured configuration restricts which image digests the pod may pull. Because the policy file is part of the launch digest, attestation transitively proves that the pod can only run the images the policy permits. This gives the strongest hardware guarantee but requires rebuilding the measured configuration each time the allow-list changes, which scales poorly across many short-lived workloads.

\item \textbf{Per-customer workload signing.} A customer's public key is baked into the pod's measured configuration, and the customer signs their image digests out-of-band (for example with cosign~\cite{ref:cosign} or notation). The in-pod agent verifies every image pull against the measured key before accepting the image. Attestation then proves that the pod will only run images the customer authorized, without requiring a new measured configuration per workload. This is the default for multi-tenant platform deployments. Customers onboard by registering a key, and from that point on they can deploy signed workloads without further platform involvement.

\item \textbf{Encrypted container images.} Images are encrypted at rest, and the decryption key is released through the CDS only to pods whose attestation matches the key's release policy. This is complementary to the two mechanisms above. It protects the image's contents against unauthorized environments rather than proving to a Verifier which image is running, and is the right choice when image confidentiality itself is the concern.
\end{enumerate}

Each of the three mechanisms anchors on something that is measured at pod launch, whether an allow-list of digests, a customer signing key, or a CDS decryption key. What differs is where the policy decision lives, which determines operational cost and who is in the loop.

\begin{figure}[H]
\centering
\includegraphics[width=0.95\linewidth]{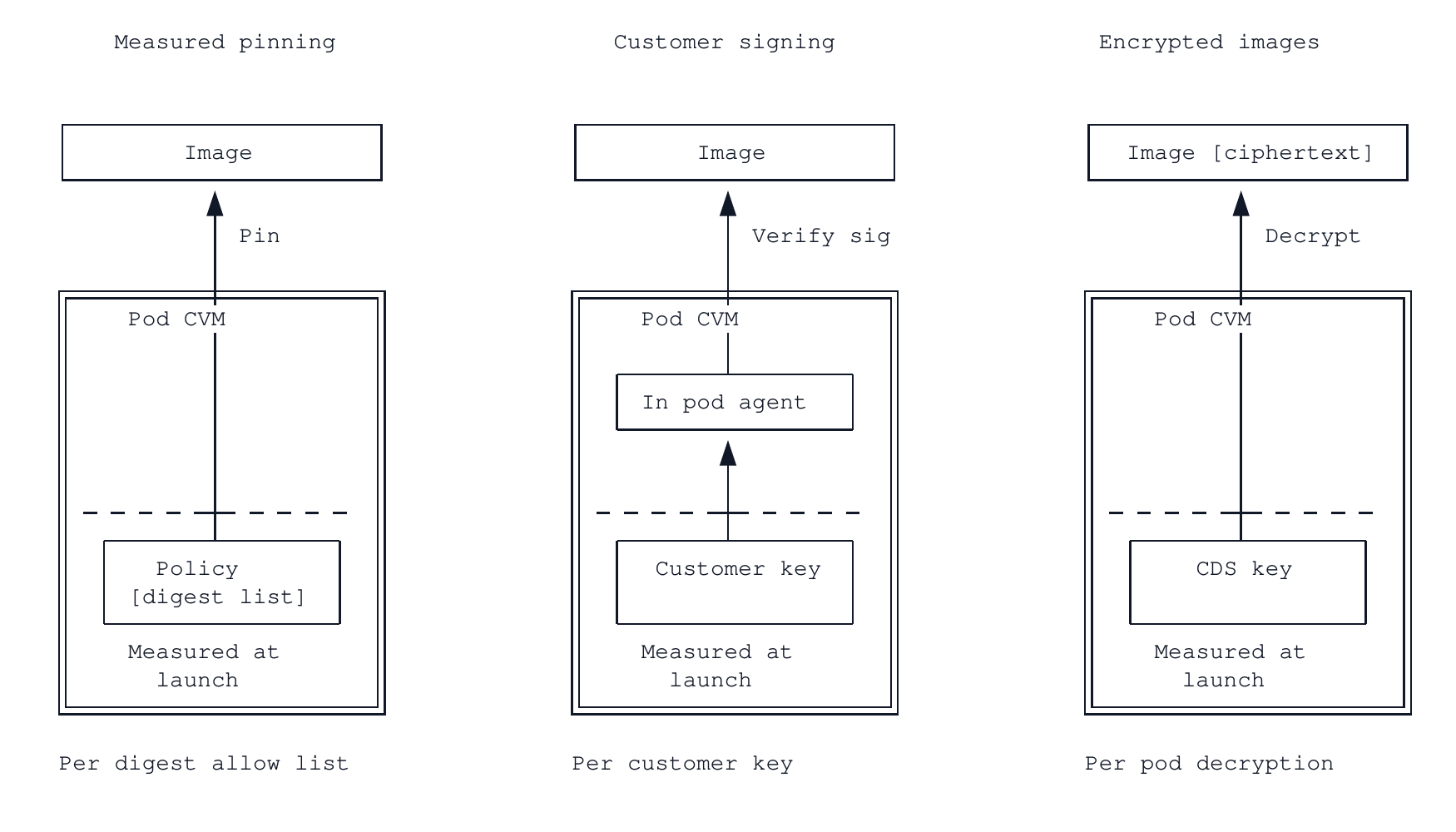}
\caption{Three image-gap mechanisms. All three anchor on something inside the pod's launch digest; they differ in whether the measured anchor is an allow-list, a signing key, or a decryption key.}
\label{fig:image-gap}
\end{figure}

\subsection{Node-level CVM}
\label{sec:node-cvm}

When the trust boundary is drawn around the worker node rather than the pod, each Kubernetes node runs as a CVM. At boot, the hardware's secure processor computes a launch measurement covering the firmware, kernel, and initial ramdisk. When GPU CC mode is available, the measurement additionally attests the GPU hardware and firmware, producing a combined report. This measurement is the foundation of the attestation chain for every pod that lands on the node.

\begin{figure}[H]
\centering
\includegraphics[width=0.85\linewidth]{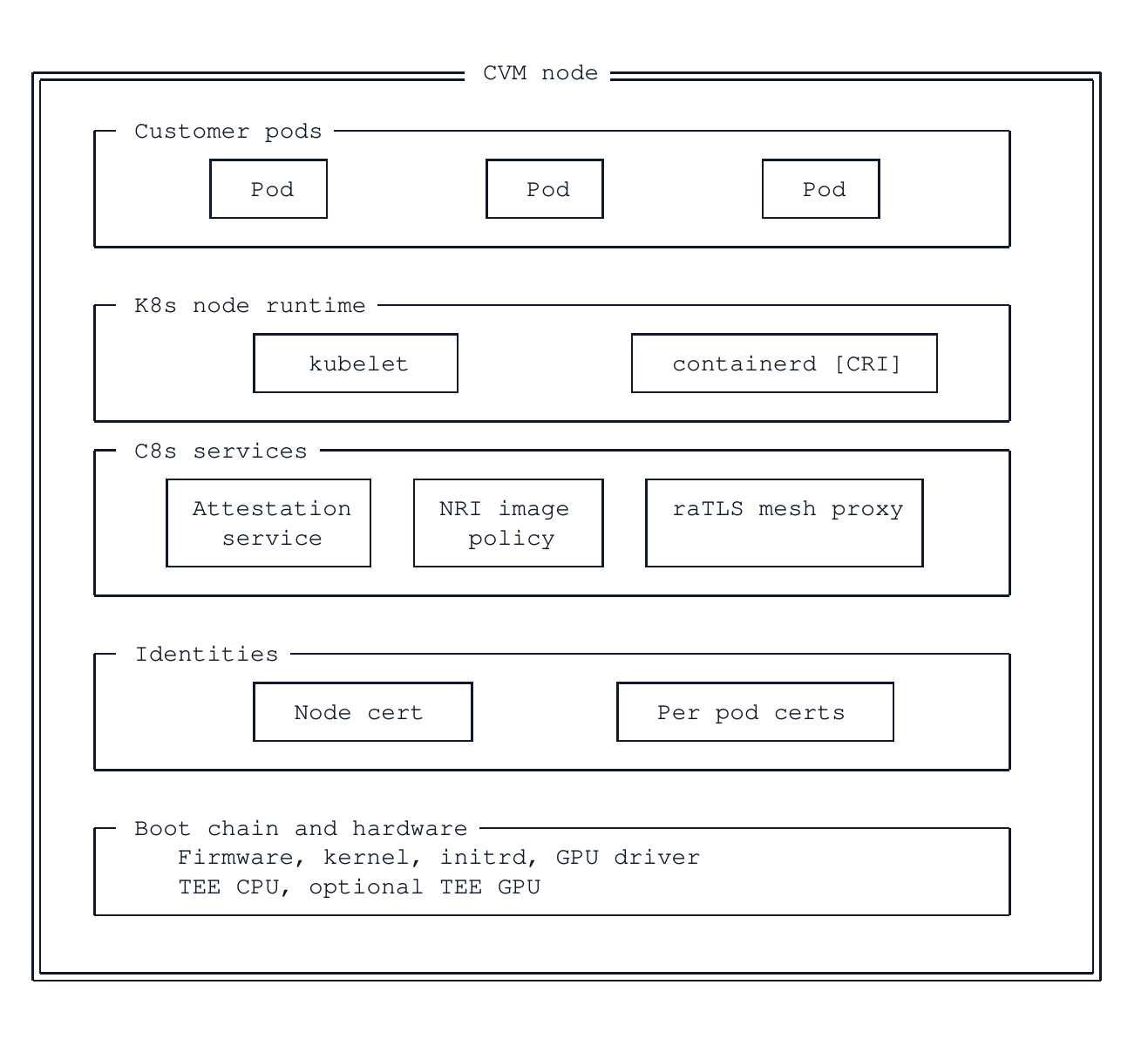}
\caption{Node-level CVM anatomy. The entire worker node boots as a confidential VM; customer pods, the Kubernetes runtime, and the C8s services (attestation, NRI policy, mesh proxy) all share the node's hardware-encrypted memory.}
\label{fig:node-cvm}
\end{figure}

Applying the design formula from \S\ref{sec:design-formula}:

\begin{enumerate}
\item The node runs as a CVM; the TEE encrypts VM memory with keys the hypervisor never possesses.
\item The secure processor computes a launch measurement at boot.
\item The CDS verifies the node's attestation report and issues a node-scoped credential. For every pod that subsequently launches on the node, the CDS issues a per-pod certificate based on composed evidence (the node's TEE report plus the pod's workload digest, signed by the node's attestation service), so each pod still holds its own attestation-rooted identity.
\item Other components verify the pod's attestation-rooted identity before accepting connections.
\item Traffic leaving any pod is encrypted via the raTLS mesh.
\end{enumerate}

All workloads running inside the CVM inherit its memory encryption. Containers within a single CVM share the node's hardware protection boundary and are separated by standard Linux kernel isolation mechanisms (namespaces, cgroups, seccomp). This is process-level isolation, not hardware isolation. Pods on the same CVM node can access each other's memory through a kernel exploit or misconfigured security context, just as they could on a standard node.

For deployments requiring hard multi-tenancy, where different customers' workloads must be hardware-isolated from one another, each tenant requires dedicated CVM nodes. The C8s architecture supports this naturally. New CVM nodes are provisioned on demand through standard Kubernetes autoscaling, and each boots through the same attestation flow. Sovereign-cloud and high-compliance deployments should treat node-level isolation as the tenancy boundary and schedule at most one tenant per CVM node.

A tighter hardware boundary than the node may be needed for per-pod hardware isolation, per-pod attestation independent of node posture, or compatibility with managed Kubernetes clusters that do not expose a confidential node pool. In that case the same design formula is applied one level down, with each pod running as its own confidential VM (\S\ref{sec:pod-cvm}). \S\ref{sec:choosing-boundary} discusses when to choose each.

\subsection{Deployment Topologies}

On a self-hosted confidential host, per-pod CVMs are launched directly by the pod runtime on the node. Each pod gets its own measured VM; the host is responsible only for scheduling and for providing the runtime.

On a managed Kubernetes cluster that does not expose a confidential node pool, per-pod CVMs are provisioned through a peer-pod path. The pod runtime delegates VM lifecycle to the cloud provider's API and each pod becomes a dedicated confidential cloud VM (for example Azure Confidential VM, AWS Nitro-based SNP instances, or GCP Confidential VM). The cluster's Kubernetes nodes remain ordinary VMs whose role is reduced to control surface (scheduling, networking, and proxying pod lifecycle calls to the cloud), which is what makes this path compatible with managed services that have no CVM node pool at all. The trade-off is that every pod launch now includes cloud VM provisioning on the cold path, which adds latency and bills per VM rather than per container; warm pools amortize this at the cost of idle capacity.

Hybrid clusters, in which some workloads use node-level CVMs and others use per-pod CVMs on the same control plane, are supported. The CDS accepts both evidence shapes.

\subsection{GPU Attachment}

GPU CC mode binds a GPU's protected state to a single confidential guest. When the boundary is at the pod, any GPU the pod uses must be attached to the pod's CVM at launch, and it cannot be multiplexed across pods while that pod is running. Per-pod deployments therefore allocate whole GPUs (or whole GPU partitions, where the hardware supports partitioning) per pod.

This is a real constraint for GPU-dense workloads. When the boundary is at the node, the GPU attaches to the node's CVM and several pods on that node can share GPU capacity, all running inside the node's trust boundary. The same sharing is not available at the pod level, because the sharing mechanisms assume multiple tenants inside the GPU's single attested context, and per-pod CVMs each want their own attested context. For workload pools that rely on cross-pod GPU sharing to reach high utilization, this consideration alone often decides where the boundary should sit.

\begin{figure}[H]
\centering
\includegraphics[width=0.92\linewidth]{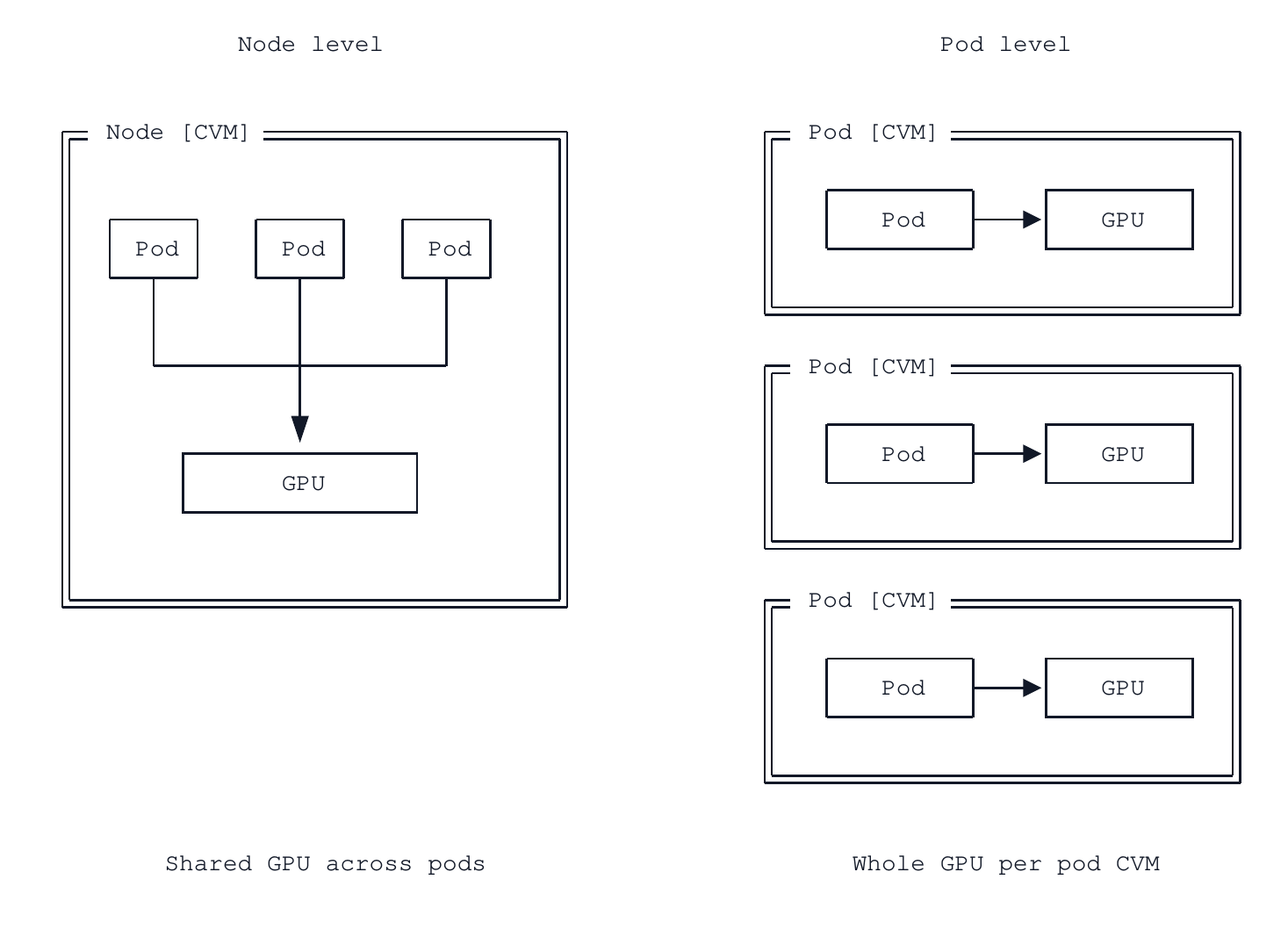}
\caption{GPU attachment comparison. Node-level deployments can time-slice or partition a single attached GPU across pods; pod-level deployments dedicate a whole GPU (or whole partition) to each pod's CVM because CC mode binds the protected context to one guest.}
\label{fig:gpu-attach}
\end{figure}

\subsection{Choosing Where to Draw the Boundary}
\label{sec:choosing-boundary}

The choice between a node-level and a pod-level boundary is a deployment-time configuration, not two separate products. The same CDS, the same raTLS mesh, the same key-release machinery, and the same client-side encryption apply in either case; what changes is the granularity of hardware isolation, the shape of attestation evidence, and a handful of operational consequences that follow from where the CVM actually lives.

In practice the pod-level boundary is the C8s default and the appropriate choice for per-pod tenancy, for managed Kubernetes deployments that do not offer a confidential node pool, and for workloads that need per-pod attestation independent of the node's posture. The node-level boundary remains attractive for single-tenant high-throughput pools (inference, training, batch analytics) where per-pod hardware isolation is unnecessary and sharing GPU capacity across pods on the same node matters for utilization. The second-order consequence is worth stating explicitly. At the pod level, the worker node stops being a confidential VM in any meaningful sense and becomes a control surface that proxies pod lifecycle to per-pod CVMs. That shift reshapes what a ``node'' is operationally. Peer-pod and virtual-kubelet-style topologies become natural, node-level GPU sharing becomes unavailable, and per-pod VM provisioning latency becomes a real cost to budget for.

The following side-by-side shows where each piece of the platform sits under the two configurations. The node-level view puts the CVM wall around the node; the pod-level view collapses the per-node DaemonSets into each pod's own measured configuration and reduces the host to a control surface.

\begin{figure}[H]
\centering
\includegraphics[width=\linewidth]{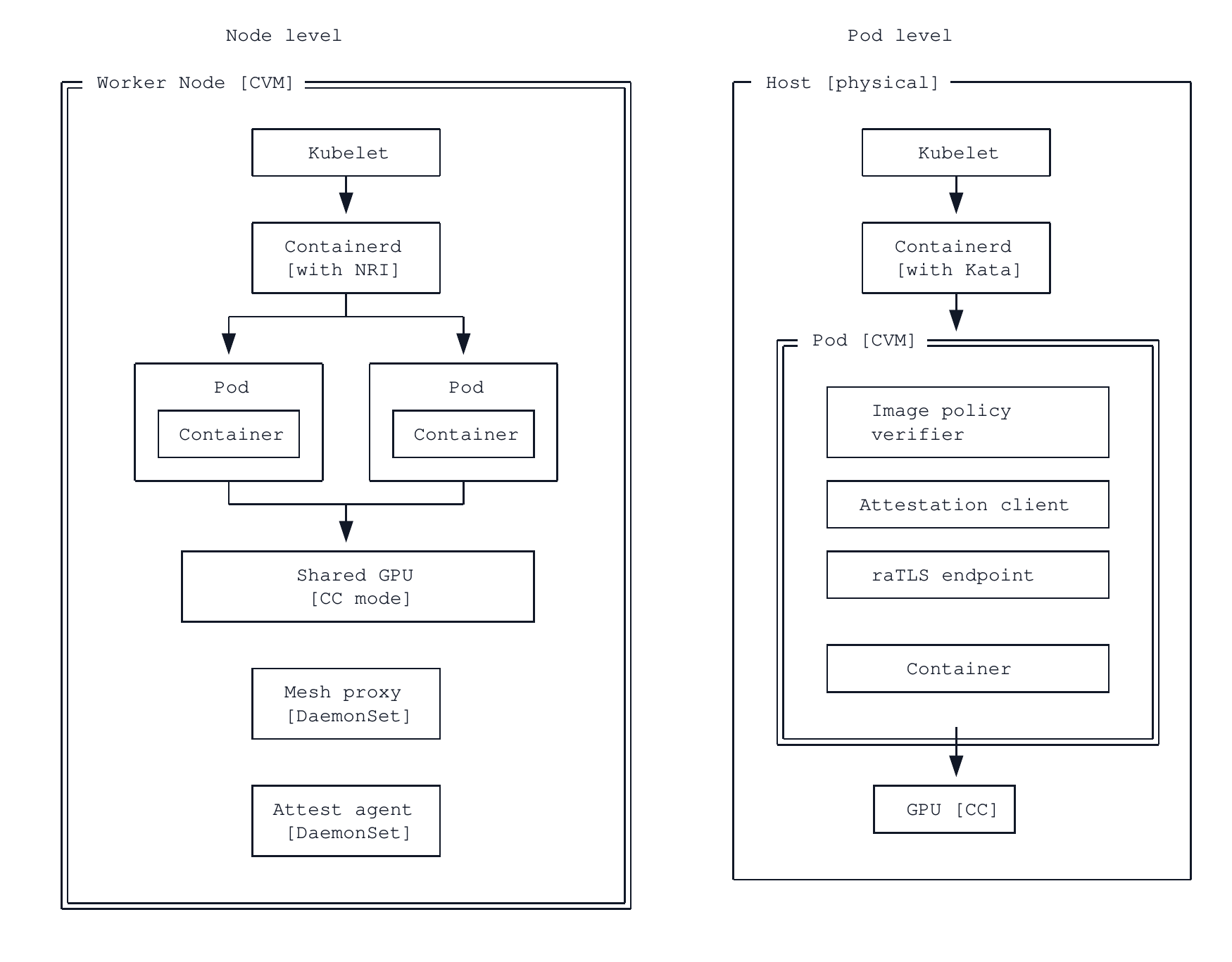}
\caption{Node-level vs.\ pod-level boundary. Same CDS, same mesh, same key-release machinery on both sides; what moves is where the CVM wall sits and which components end up inside it.}
\label{fig:node-vs-pod}
\end{figure}

The trade-offs in summary:

\begin{table}[H]
\centering
\footnotesize
\renewcommand{\arraystretch}{1.15}
\begin{tabularx}{\linewidth}{@{} >{\bfseries}p{0.18\linewidth} X X @{}}
\toprule
Axis & Node-level boundary & Pod-level boundary \\
\midrule
Isolation boundary & Node. Pods on the same node share kernel-level isolation. & Pod. Each pod is its own confidential VM with hardware-isolated memory. \\
Attestation granularity & Composed: node attestation plus per-pod workload digest. & Direct: per-pod attestation bound to the pod's launch digest. \\
Image and workload gating & Host-side NRI check against a signed allow-list before containers start. & In-pod measured policy, per-customer workload signing, or both, enforced by the in-pod agent before the image pull completes. \\
Container-image proof & Image digest checked against the signed manifest at launch, on the host. & Image is not in the launch digest; proof comes from the in-pod signing or policy check, which itself is in the digest. \\
Managed K8s compatibility & Works directly on confidential node pools where the cloud offers them (AKS, EKS, GKE with confidential nodes). & Works anywhere, via peer-pod-style provisioning of confidential cloud VMs per pod, including clusters whose nodes have no TEE support. \\
Startup latency & Pod scheduling plus container pull, on the order of seconds. & Bare-metal per-pod CVMs add a VM boot to the pod's cold path; peer-pod deployments additionally incur cloud VM provisioning (tens of seconds) per pod, which warm pools can amortize at the cost of idle capacity. \\
GPU attachment & GPU attaches to the node's CVM and is shared across pods on that node through standard GPU-sharing mechanisms (time-slicing, MPS, MIG) inside the node's trust boundary. & GPUs are assigned to a pod's CVM at launch and cannot be shared across pods while the pod runs, because GPU CC mode binds the protected context to a single guest. Whole GPUs or whole GPU partitions are allocated per pod. \\
Per-pod resource overhead & Low. Pods run as regular processes and share the node's guest kernel and agent stack. & Meaningful. Each pod carries its own guest kernel, in-pod agent, attestation agent, and policy enforcement, typically on the order of a few hundred megabytes of memory before the workload itself. \\
Tenancy model & Hard multi-tenancy requires dedicated nodes per tenant. Within a single tenant, pods share the node's trust boundary. & Hard per-pod isolation by construction. Natural fit for multi-tenant clusters without dedicated tenant nodes. \\
Operational complexity & Cluster looks like standard Kubernetes with a confidential node image, a signed image policy manifest, and the mesh DaemonSets. An ops team already running Kubernetes has to add one node image and one allow-list. & Cluster needs per-pod CVM lifecycle management (measured configurations per workload class or per customer), a signed-workload enrollment flow, and, on managed Kubernetes, the peer-pod provisioning path with its own cloud-API integration, quotas, and failure modes. Operationally this is closer to running a per-pod VM fleet than a Kubernetes pod fleet. \\
Worker-node role & The worker node is a confidential VM and carries real workload trust. & The worker node is a control surface and holds no workload trust, because the workload's TEE is the pod itself. In the limit, the node can be a plain VM. \\
\bottomrule
\end{tabularx}
\caption{Trade-offs between the two boundary placements.}
\end{table}

Mixing boundaries within a single cluster is supported, so node-level CVMs for throughput-bound pools and pod-level CVMs for tenant-isolated workloads can coexist under the same CDS. Further refinement (pod-level raTLS endpoints, per-workload policy caching, and the roll-out of the per-customer signing model) is discussed in \S\ref{sec:open}.

\subsection{Certificate Distribution Service (CDS)}
\label{sec:cds}

The CDS is the Verifier and the root of the certificate chain within the trust boundary. It sits in the RATS \emph{background-check} role~\cite{ref:rats}: pods produce Evidence, the CDS appraises it, and mesh peers act as Relying Parties that trust the CDS's prior appraisal via the issued certificate. It runs as a set of attested pods and performs four functions:

\begin{enumerate}
\item \textbf{Attestation verification.} When an unattested pod requests attestation, the CDS issues it a challenge nonce and the Attester responds with an attestation report. The CDS verifies that: (a) the endorsement key chains to the hardware manufacturer's root of trust, (b) the report signature is valid, (c) the launch measurement matches a reference value in the allow-list, (d) the Trusted Computing Base (TCB) version meets minimum requirements, and (e) the signed attestation nonce matches the expected value.
\item \textbf{Certificate issuance.} Upon successful appraisal, the CDS issues a raTLS mesh certificate to the pod for use in inter-pod traffic. The CDS does not issue externally facing certificates; the ingress router uses a public-CA-issued certificate bound to a TEE-internal key, paired with a CDS-countersigned freshness beacon that clients verify (see \S\ref{sec:default-ingress} and \S\ref{sec:default-tls}).
\item \textbf{Manifest signing.} The CDS signs the image policy manifest consumed by the image policy enforcement layer (\S\ref{sec:nri}); the manifest carries the reference values appraisal uses. Relying Parties verify it via standard certificate chain validation.
\item \textbf{Key brokering.} The CDS releases attestation-gated secrets to attested pods whose attestation reports match per-key release policy. The full release flow is described in \S\ref{sec:key-broker}.
\end{enumerate}

These responsibilities are summarized in the following figure; each subsequent subsection drills into deployment, state management, and the minimal-key property.

\begin{figure}[H]
\centering
\includegraphics[width=0.92\linewidth]{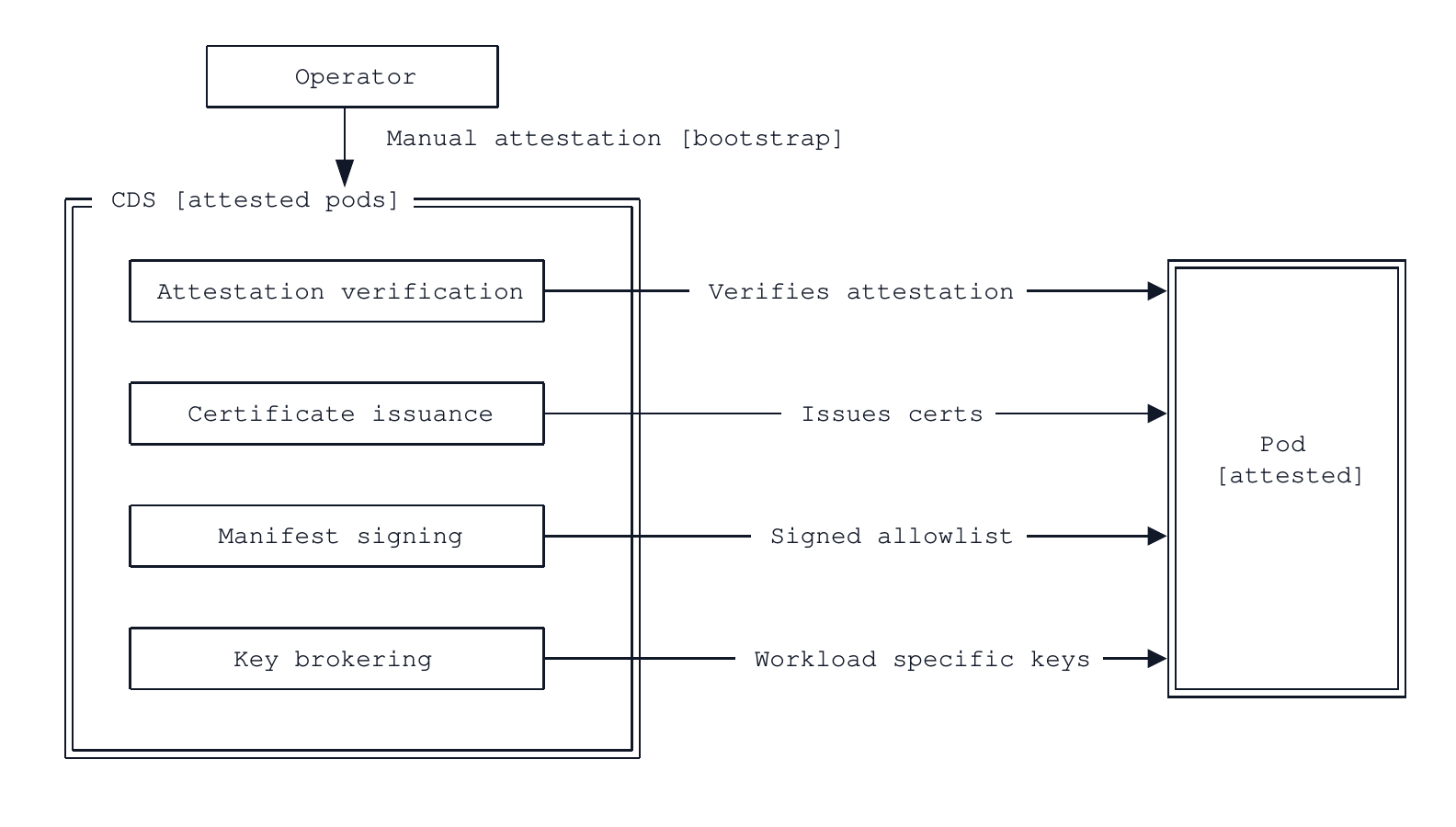}
\caption{CDS responsibilities. The CDS is itself a CVM manually attested by the operator at bootstrap; from there, every outgoing channel is anchored in an attestation decision.}
\label{fig:cds-services}
\end{figure}

\subsubsection{Deployment Model}
\label{sec:cds-deploy}

The CDS is deployed as an active/active pair so it does not become a single point of failure. Each replica runs as an attested pod with the same measurement and serves traffic independently. A CA-style signing operation is idempotent given the same inputs (the operator-signed allow-list and an incoming attestation report), so no consensus protocol is needed between replicas. All replicas share the same CDS measurement, signing key, and identity, so the operator's one-time bootstrap covers the set rather than each replica individually.

The signing key is generated inside the first replica's CVM and never leaves CVM-encrypted memory in plaintext. When a new replica starts, it presents an attestation report binding a CVM-resident public key to the CDS measurement; an existing replica verifies the report against its own measurement and wraps the signing key to that public key (the same primitive used for application secrets in \S\ref{sec:key-broker}). The control plane handles only ciphertext wrapped to a key it cannot reach. If every replica becomes unavailable simultaneously, the in-memory signing key is lost; recovery is a re-bootstrap, in which the operator re-attests a fresh CDS, a new signing key is generated, and mesh certificates reissue under the new key as pods re-attest. The allow-list is operator-signed, so each replica independently verifies the signature on every update, and updates are pushed to every replica.

The CDS exposes an attested endpoint that returns the currently active allow-list and the immediately previous version. Because the endpoint is served by the measured CDS image inside its CVM, a verifier reading from this endpoint receives the allow-list state the CDS is actually enforcing rather than what the operator claims it is. This narrows the gap between policy-as-deployed and policy-as-claimed without requiring the operator's trust. A fuller append-only history with signed consistency roots over the entire allow-list sequence is treated as future work in \S\ref{sec:open}.

Kubernetes ConfigMaps and Secrets, including those managed by SealedSecrets controllers, External Secrets Operator, or similar tooling, are unsuitable for CDS state because the kube-apiserver always has access to plaintext. These mechanisms protect against certain attacks (stolen etcd disks, leaked Git repositories) but do not protect against a compromised control plane, which is explicitly part of our threat model. Confidentiality against the control plane requires that decryption happen inside TEE-encrypted memory, with keys derived from hardware attestation, so that no Kubernetes component ever handles plaintext.

Other infrastructure services that exist in node-level deployments (the pod attestation agent, the raTLS mesh interceptors, and the NRI image policy enforcers) run as DaemonSets and attest to the CDS on startup like any other pod, presenting composed evidence (the node's TEE report plus their own workload measurement) and receiving per-pod certificates through the standard automated flow. Where the boundary is drawn at the pod (the default), there are no DaemonSet attestation agents to bootstrap. Each pod is its own CVM and presents its own TEE report directly to the CDS at start, and the equivalent gating relocates into the pod's measured configuration (\S\ref{sec:pod-cvm}).

\subsubsection{Bootstrapping and Trust Anchoring}
\label{sec:cds-bootstrap}

The CDS is the first component to start in a fresh cluster, because every other attested pod needs the CDS to issue its certificate. Someone has to be trusted first; in C8s that someone is the operator, and the decision is bounded to a one-time manual attestation event that proceeds as follows:

\begin{enumerate}
\item The operator installs the CDS via its Helm chart or equivalent.
\item Kubernetes schedules the CDS replicas, each as an attested pod running the CDS's measured image.
\item The operator retrieves attestation evidence from one replica and verifies it against the hardware manufacturer's root of trust. This verification establishes that the CDS measurement is running on genuine hardware, and that the measurement matches the expected CDS image digest.
\item Key material for management of the allow-list is bootstrapped by the cluster owner to ensure integrity of the allow-list and thus images permitted by image-policy enforcement.
\item The operator signs the initial allow-list and uploads it to the CDS.
\item Because every CDS replica is an attested pod with the same measurement, the operator's verification applies to all replicas at this measurement. Replacing the CDS image (e.g., upgrading versions) produces a new measurement and requires re-attestation.
\end{enumerate}

This trust anchoring is the architecture's one human-in-the-loop step. Every subsequent attestation in the cluster chains to the operator's verification of the CDS measurement.

\begin{figure}[H]
\centering
\includegraphics[width=0.65\linewidth]{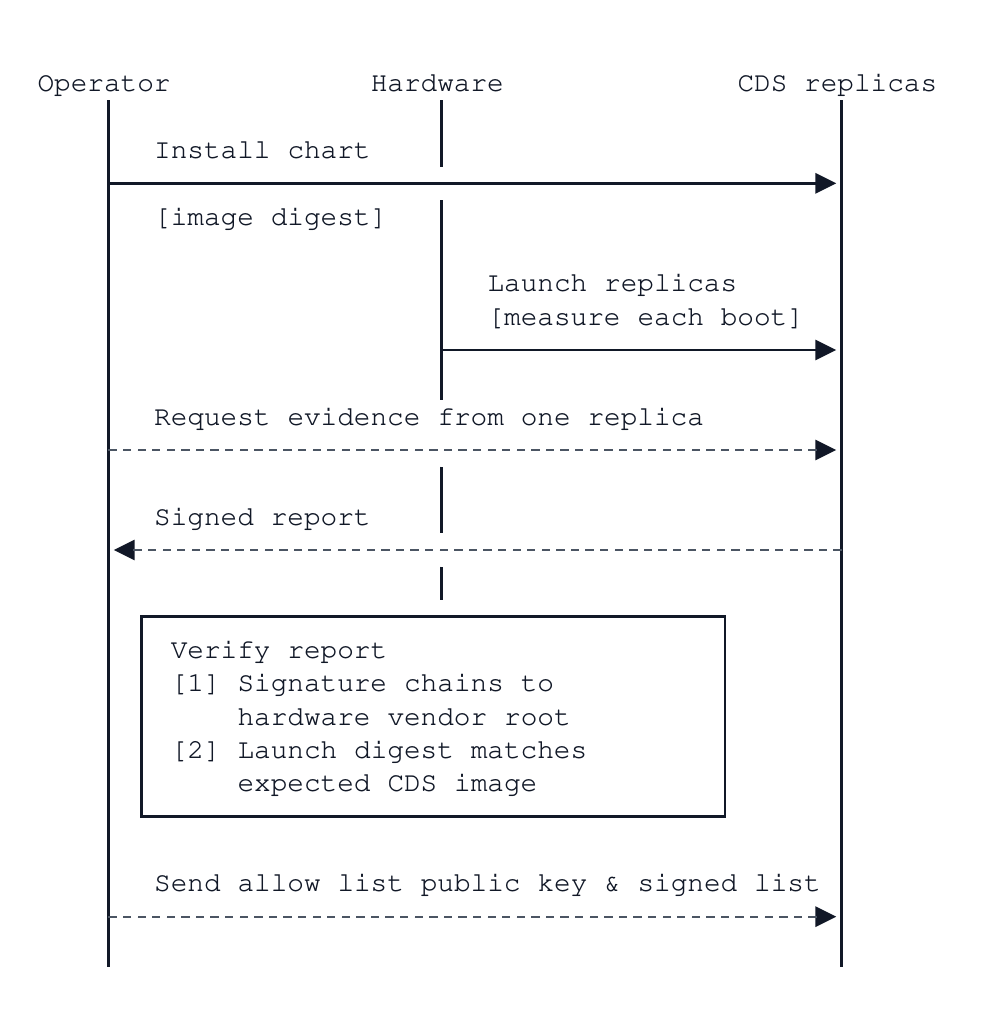}
\caption{CDS bootstrapping. The operator attests the CDS measurement once; this covers every replica at that measurement, present and future.}
\label{fig:cds-bootstrap}
\end{figure}

\textbf{Consequences of unavailability.} If the CDS becomes unavailable, existing certificates keep working until they expire. New attested pods cannot obtain certificates, so new pods cannot join the mesh. Running workloads are unaffected for the lifetime of their existing certificates. This bounds the blast radius of CDS downtime to ``can't onboard new pods'' rather than ``cluster goes dark.''

\subsubsection{Key Minimization}

An important property of the CDS is what it does not hold. The CDS possesses its signing key, which it uses to issue certificates and to sign freshness beacons (\S\ref{sec:default-tls}). It does not hold the raTLS session keys that protect data in transit, the memory-encryption keys used by any CVM, or any application-layer decryption key outside the release flow described in \S\ref{sec:key-broker}. Those keys live inside the boundary of CVMs and never leave them. Session keys are ephemeral and negotiated per connection by the endpoints. CVM memory keys never leave the hardware secure processor. Brokered application secrets are forwarded to the attested pod that requested them and held only in that pod's hardware-encrypted memory.

This scoping narrows the blast radius of a CDS compromise. An adversary who obtains the CDS signing key can issue fraudulent certificates going forward and, at worst, impersonate attested workloads to peers that accept CDS-issued identity. They cannot decrypt past traffic retroactively, break in-flight sessions protected by ephemeral keys, or read CVM memory whose keys the CDS has never possessed. The CDS's core responsibility is a certificate authority and attestation-gated broker. Application-layer keys flow through the brokering protocol in \S\ref{sec:key-broker}, which supports two modes. In neither mode does the CDS hold application keys at rest. The recommended wrapped-brokering mode has the CDS handle only ciphertext, never seeing plaintext keys at all. The fallback direct-brokering mode has the CDS proxy plaintext keys from a customer KMS to attested pods, with each key exposed only in the CDS's CVM memory for the duration of its proxy hop. A CDS compromise active during a direct-mode hop can read the key in flight, and an adversary persisting in the CDS will see every subsequent direct-mode key until detection. No keys are retroactively recoverable from past completed hops.

\textbf{Certificate rotation is re-attestation.} One further implication follows from this scoping. Because the CDS has no long-lived secret corresponding to any pod's mesh identity beyond the outstanding certificate itself, rotating a pod's certificate is inseparable from re-attesting the pod. The CDS re-verifies the pod's current attestation report, re-checks the measurement against the allow-list, and re-checks any key-binding fields before issuing a replacement cert. Renewal fails if the pod's TCB has fallen below the minimum or its measurement has been revoked, and the pod drops out of the mesh at the old certificate's expiry. Expiry plus re-attestation is the revocation mechanism; there is no separate revocation channel to operate or to fail. \S\ref{sec:freshness} details how short-lived certificates and key-digest binding make this work without per-handshake attestation overhead.

\subsubsection{Attestation-gated Key Brokering}
\label{sec:key-broker}

Beyond certificate issuance, the CDS is the broker for application-layer secrets whose release should be tied to a specific attested measurement. The pattern fits any artifact whose release policy can be expressed in terms of measurement, workload identity, and per-key release rules: encrypted datasets, signing keys for downstream services, decryption keys for proprietary code, and model weights are all instances of the same shape. C8s supports two brokering modes. Wrapped brokering is the default and recommended mode and is described first. Direct brokering is a fallback for customers whose key-management workflow cannot supply a deposit service.

\textbf{Release decision.} Both modes share the same release decision. When an attested workload requests a secret, the CDS evaluates three conditions before authorizing release.

\begin{enumerate}
\item \textbf{Substrate attestation.} The CVM's launch measurement is on the allow-list, and the firmware or TCB version meets the minimum.
\item \textbf{Workload authorization.} The container image digest matches the signed policy manifest (enforced by the NRI image policy enforcer at the node, or by the in-pod measured policy at the pod).
\item \textbf{Per-key policy.} The specific secret is authorized for release to this measurement-and-digest combination.
\end{enumerate}

Only when all three are satisfied does the CDS proceed with release. What ``release'' means differs between the two modes.

\textbf{Wrapped brokering (default).} The customer runs a key deposit service that holds the plaintext secret. At pod startup, the attested pod generates an ephemeral keypair inside its CVM and presents an attestation report binding its public key to the measurement, using the same key-binding mechanism \S\ref{sec:freshness} uses for raTLS certificates. The CDS verifies the three release conditions above and forwards the request, together with the pod's attestation report and bound public key, to the customer's deposit service. The deposit service independently verifies the attestation against the hardware vendor's root of trust, wraps the secret to the pod's public key (for example with HPKE~\cite{ref:hpke}), and returns the ciphertext. The CDS forwards the ciphertext to the pod over the existing raTLS connection. The pod unwraps the secret inside its CVM-encrypted memory.

In this mode, the CDS handles ciphertext only. It never holds the plaintext secret, and a CDS compromise does not yield secret plaintext. The customer remains in the loop on every release and can deny releases their deposit service does not approve. Two onboarding paths sit on top of this mode, illustrated here for the canonical case of encrypted model weights. In the \emph{platform-managed} path, the artifact owner uploads plaintext and the platform encrypts on ingest, leaving the owner's workflow unchanged. In the \emph{provider-managed} path, the artifact owner encrypts before upload and operates the deposit service themselves, retaining direct control over both the encryption and the release decision. The same two paths apply to other artifact classes (datasets, code, downstream credentials).

\textbf{Direct brokering.} Where a customer cannot supply a deposit service that wraps keys to per-pod public keys, the CDS instead acts as an attested proxy to the customer's existing KMS. The CDS holds no keys at rest in this mode. On a matching request, the CDS authenticates to the customer's KMS by presenting its own attestation report; the customer's KMS attests the CDS against the hardware vendor's root of trust and releases the plaintext key over an authenticated channel. The CDS forwards the plaintext key to the requesting pod over the raTLS connection. The key transits the CDS's CVM-encrypted memory only for the duration of the proxy hop and is never persisted within the CDS.

This mode is appropriate when the customer's KMS supports ``release to an authenticated client'' but not ``wrap to an arbitrary external public key,'' which is the common case for cloud-native KMS products. The trade-off relative to wrapped brokering is exposure during the proxy hop.

In either mode, the released material reaches only an attested pod over a raTLS connection, never persists outside CVM-encrypted memory, and is rotated by re-attestation rather than by a separate revocation channel.

\begin{figure}[H]
\centering
\includegraphics[width=0.92\linewidth]{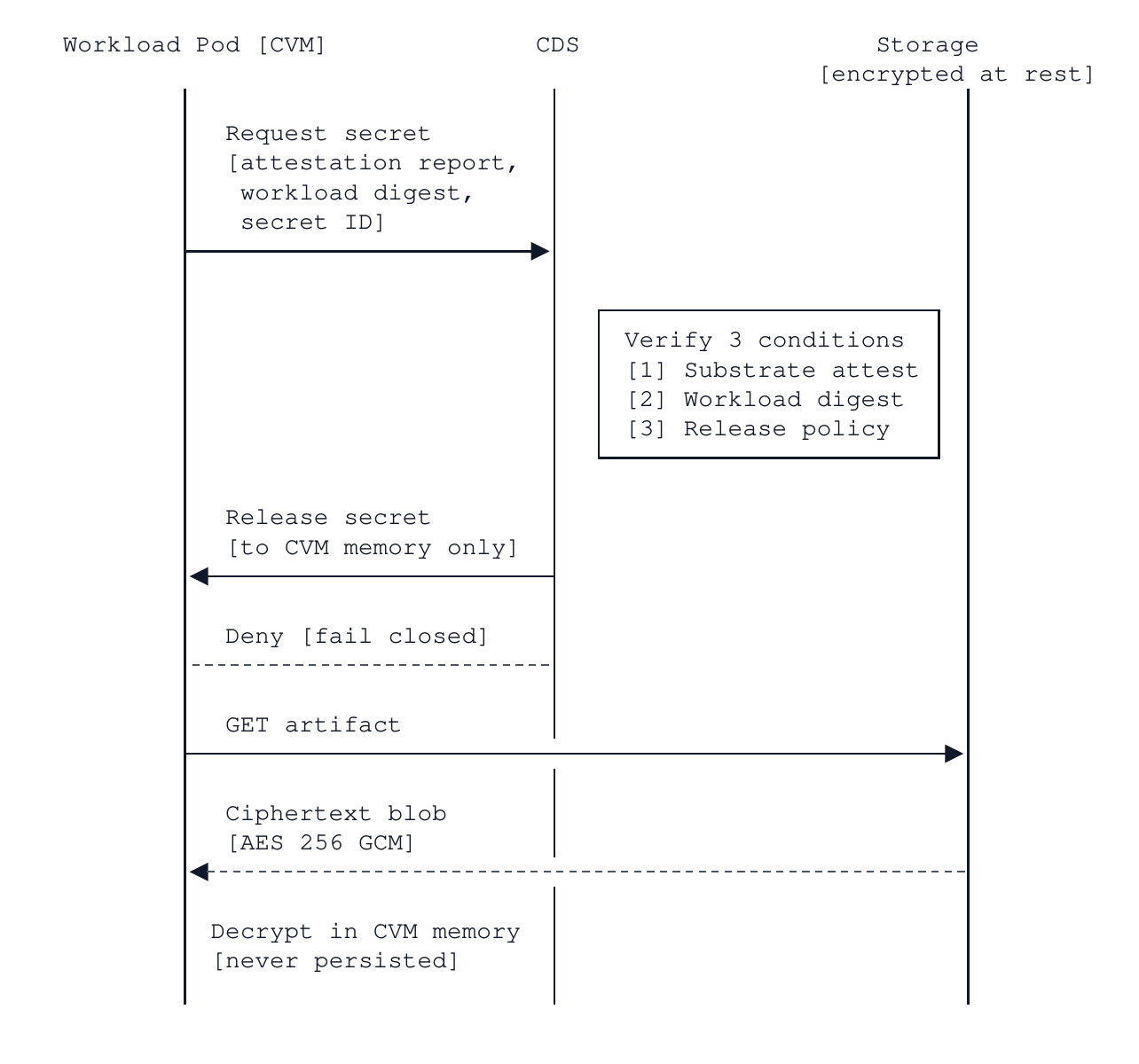}
\caption{Attestation-gated key delivery. The CDS releases a secret only when all three release conditions match; the attested pod fetches ciphertext from untrusted storage and decrypts it exclusively inside CVM-encrypted memory.}
\label{fig:key-brokering}
\end{figure}

\textbf{Multi-key isolation.} Each secret has its own release policy, and the CDS ties release to specific workload measurements rather than to which node or pod hosts the workload. Two providers' secrets can coexist on shared physical hardware. Provider A's secret is never released to provider B's workload, even on the same node, because the workload digests differ.

\textbf{Fit and limits.} The CDS is deliberately scoped to secrets whose release can be tied to an attested measurement. Long-lived operational credentials and general-purpose application secrets are a poor fit. For those, C8s integrates with an external secrets manager such as HashiCorp Vault~\cite{ref:vault} through a proxy layer that sits inside the trust boundary: pods authenticate to the proxy with their per-pod CDS-issued identity, and the proxy brokers access to the external store. The CDS itself stays scoped to attestation-gated secrets; general secret management lives in the system organizations already use.

\begin{figure}[H]
\centering
\includegraphics[width=\linewidth]{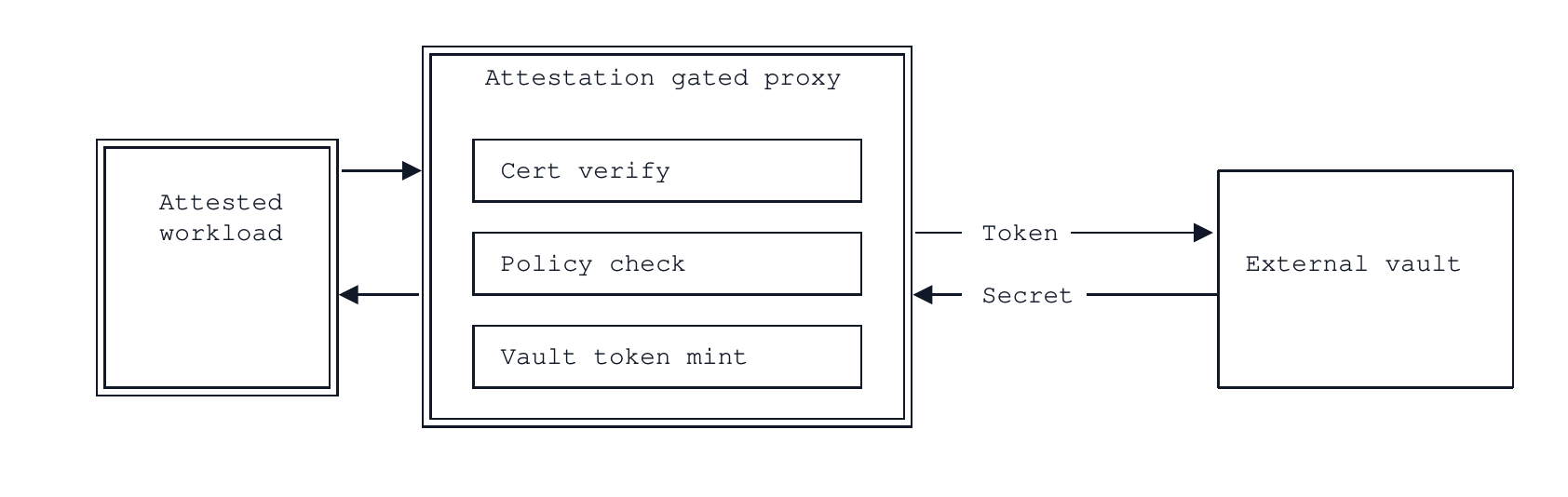}
\caption{Vault integration. The attestation-gated proxy sits inside the trust boundary; the external vault never sees a workload directly, only the proxy's identity after attestation and policy pass.}
\label{fig:vault}
\end{figure}

\subsection{NRI Image Policy Enforcer}
\label{sec:nri}

In standard Kubernetes, the control plane dictates what runs on each node. The scheduler selects a node, the API server instructs the kubelet, and the kubelet directs the container runtime to launch the container. The node has no independent authority to refuse.

This is incompatible with a trust model that excludes the control plane. A compromised or misconfigured control plane could schedule a malicious sidecar alongside the workload, one that reads decrypted secrets from shared memory or exfiltrates query data.

The NRI image policy enforcer addresses this by intercepting every container launch request at the node level. The enforcer operates as an NRI plugin, receiving container creation events from the container runtime. For each event, it extracts the image digest and checks it against a signed policy manifest. If the digest does not appear in the manifest, the container launch is rejected.

The policy manifest is signed by the CDS and carries the reference values the enforcer compares against. Digests in the manifest correspond to images built by Kettle, the attested build system, which acts as the Reference Value Provider for image identity; the operator is the Reference Value Provider for which of those digests are authorized to run. The chain of trust runs as follows. Kettle attests the build, the output digest is recorded in the operator-signed allow-list, the CDS signs the resulting manifest, and the NRI enforcer checks incoming containers against it. The control plane is excluded from this chain entirely.

The following table summarizes the attacks a compromised control plane can attempt and the mechanism that prevents each:

\begin{table}[H]
\centering
\footnotesize
\renewcommand{\arraystretch}{1.2}
\begin{tabularx}{\linewidth}{@{} L{0.34\linewidth} X @{}}
\toprule
\textbf{Attack} & \textbf{Mitigation} \\
\midrule
Schedule malicious sidecar & NRI enforcer rejects: digest not in signed policy. \\
Leak secrets via environment variables & Application-layer secrets are released by the CDS to attested pods and held only in hardware-encrypted memory, never in Kubernetes Secrets or environment variables. \\
Inject a malicious node into the cluster & Node must attest to CDS; attacker cannot produce valid attestation report. \\
Read workload traffic on the pod network & All traffic is raTLS-encrypted; control plane has no valid mesh identity. \\
Man-in-the-middle internal traffic & raTLS requires CDS-issued certificates; attacker cannot obtain valid credentials. \\
\bottomrule
\end{tabularx}
\caption{Compromised-control-plane attacks and their mitigations.}
\end{table}

The enforcement path is shown below. The NRI hook interposes between containerd and the running container, so a rejection is indistinguishable from ``container never started'' as far as the control plane is concerned. There is no partial start, no ephemeral running state, and no opportunity to leak data before the policy check completes.

\begin{figure}[H]
\centering
\includegraphics[width=0.92\linewidth]{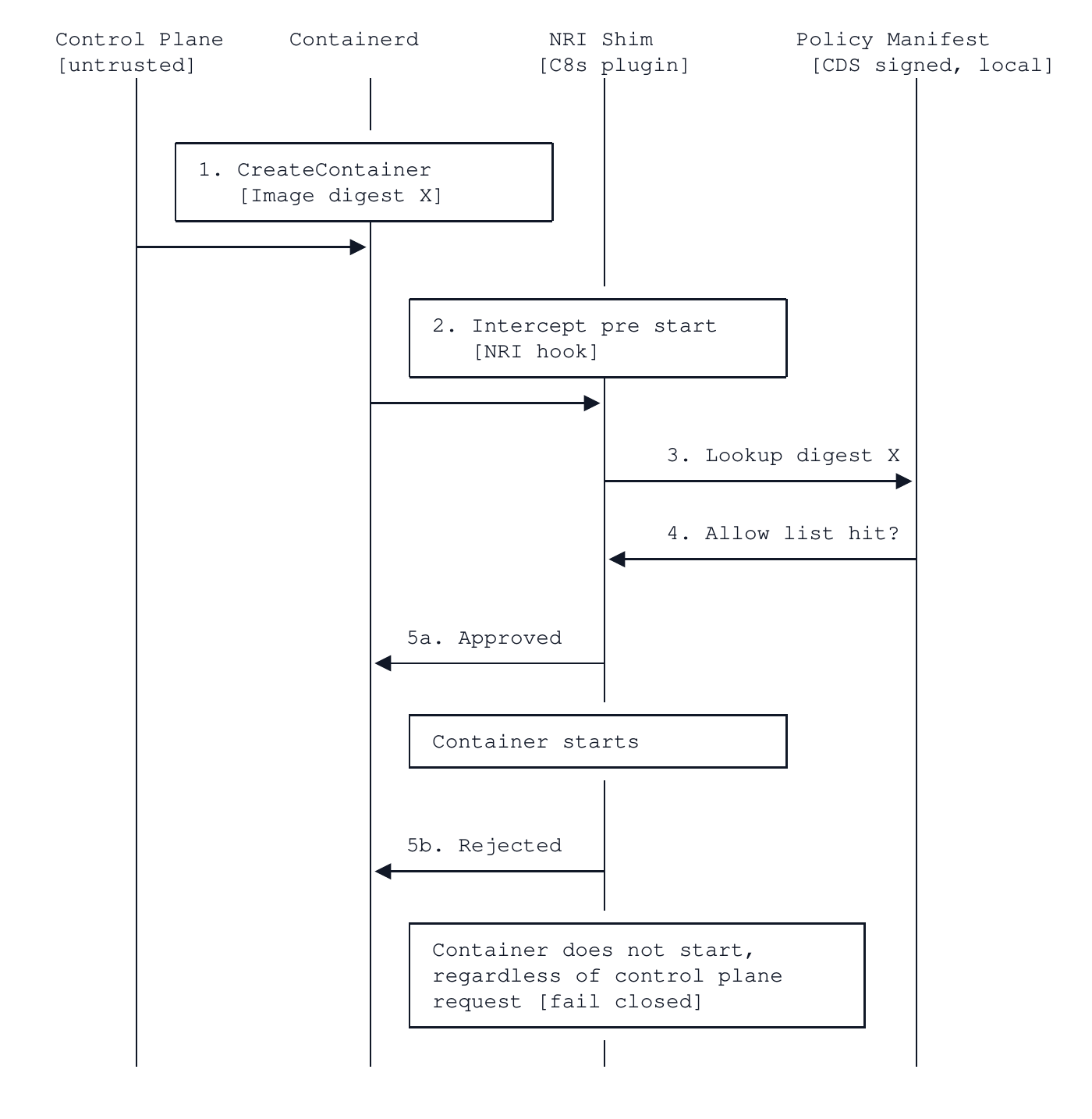}
\caption{Container startup under NRI enforcement. The policy manifest is local and CDS-signed, so the rejection decision is independent of any live control-plane interaction.}
\label{fig:nri-sequence}
\end{figure}

The NRI enforcer as described here operates at the node, which is appropriate when the node is the CVM and the host can see every container before it starts. When the trust boundary is drawn at the pod, the host cannot observe the post-boot container pull at all, and the same gating role relocates into the pod's measured configuration. A policy file or a customer key is baked into the launch digest and enforced by the in-pod agent before the image is allowed to load. The chain of trust (Kettle-attested build $\rightarrow$ signed manifest or signed workload $\rightarrow$ measured enforcement) is preserved; only the enforcement point moves.

\subsection{raTLS Mesh}
\label{sec:ratls}

Pod-to-pod traffic in Kubernetes flows through the cluster network. The CNI plugin handles routing, and traffic may traverse virtual switches, physical infrastructure, or cloud networking layers outside the operator's control. By default, this traffic is plaintext.

The network cannot be placed inside the trust boundary because the operator does not control the CNI, and network infrastructure cannot be attested. The architecture therefore applies the encrypt-before-exposure strategy via a remote attestation TLS (raTLS) mesh.

The pattern will be familiar to readers acquainted with service meshes such as Istio~\cite{ref:istio}, where each workload receives a SPIFFE~\cite{ref:spiffe} identity and a proxy terminates mTLS on its behalf. The raTLS mesh applies the same pattern with one substitution. Identity is rooted in hardware attestation via the CDS rather than in a Kubernetes-derived identity such as a service account. The certificates change meaning, not the mechanics. A certificate in a standard service mesh proves that a trusted CA issued it to a workload with a particular name. A raTLS certificate additionally proves that the workload launched on hardware-attested substrate and that its measured code matches policy.

Where the trust boundary is drawn at the node, per-pod mesh identity is produced by composed attestation. The node boots as a CVM and presents its TEE report to the CDS, receiving a node-scoped credential rooted in the hardware manufacturer chain. An attestation service (a DaemonSet whose own identity chains through the node's attestation) runs inside the CVM. When a new pod launches, the NRI path verifies its image digest against the signed policy manifest, and the attestation service produces composed evidence covering the node's TEE report, the pod's workload digest, the pod's identity (namespace, pod UID), and the pod-generated ephemeral public key, all signed by the attestation service's CDS-issued key. The pod presents this evidence to the CDS, which verifies the hardware chain, the service's signature, and the workload digest against policy, then issues a per-pod certificate bound to the pod's ephemeral key.

Where the boundary is drawn at the pod, the mechanism is simpler. The pod's own TEE report directly measures its boot state, and the pod presents that hardware report to the CDS without an intermediate agent. The CDS interface is unchanged. It receives attestation evidence, verifies the chain, checks the measurement against policy, and issues a per-pod certificate. The mesh, the certificates it issues, and the semantic guarantee read identically from the network's point of view.

Once a per-pod certificate is issued, it needs to reach the mesh proxy that will present it on the pod's behalf, and the pod's outbound traffic needs to reach that proxy. An init container in each pod handles both concerns. It requests the pod's certificate from the CDS at start, hands the certificate to the node-level mesh proxy, and arranges that the pod's TCP traffic is transparently redirected through the mesh. The workload itself is unchanged. It sends plaintext to pod-IP destinations exactly as it would on any Kubernetes cluster, and the mesh handles interception, identity presentation, and encryption. Nothing in the application code has to know about attestation, certificates, or confidential compute.

The mesh terminates at the pod, not the node. Every raTLS certificate corresponds to one pod whose evidence verified against policy at issuance time. Inter-pod traffic is intercepted at pod boundaries and encrypted end-to-end between attested pod endpoints. Pods that share a node still talk to each other over the mesh, and there is no in-node plaintext hop.

The semantic guarantee of a raTLS connection is stronger than standard mTLS. Rather than proving only that a trusted CA issued the certificate, it proves that the CDS issued this certificate to a specific pod whose workload digest matched policy, launched on a hardware-attested CVM substrate. Both ends of every internal connection hold attestation-rooted identity at pod granularity. The underlying network infrastructure sees only ciphertext between verified pod peers. An attacker who compromises the network can observe encrypted traffic but cannot participate in it, inject traffic, or decrypt it.

The following sequence shows a single pod-to-pod call under a node-level configuration with a shared mesh proxy. The workload code sends plaintext to a pod IP; interception, identity presentation, and raTLS happen at the mesh boundary. Under a pod-level configuration the mesh proxy lives inside each pod's CVM and the two hops collapse, but the on-the-wire guarantee is the same.

\begin{figure}[H]
\centering
\includegraphics[width=0.95\linewidth]{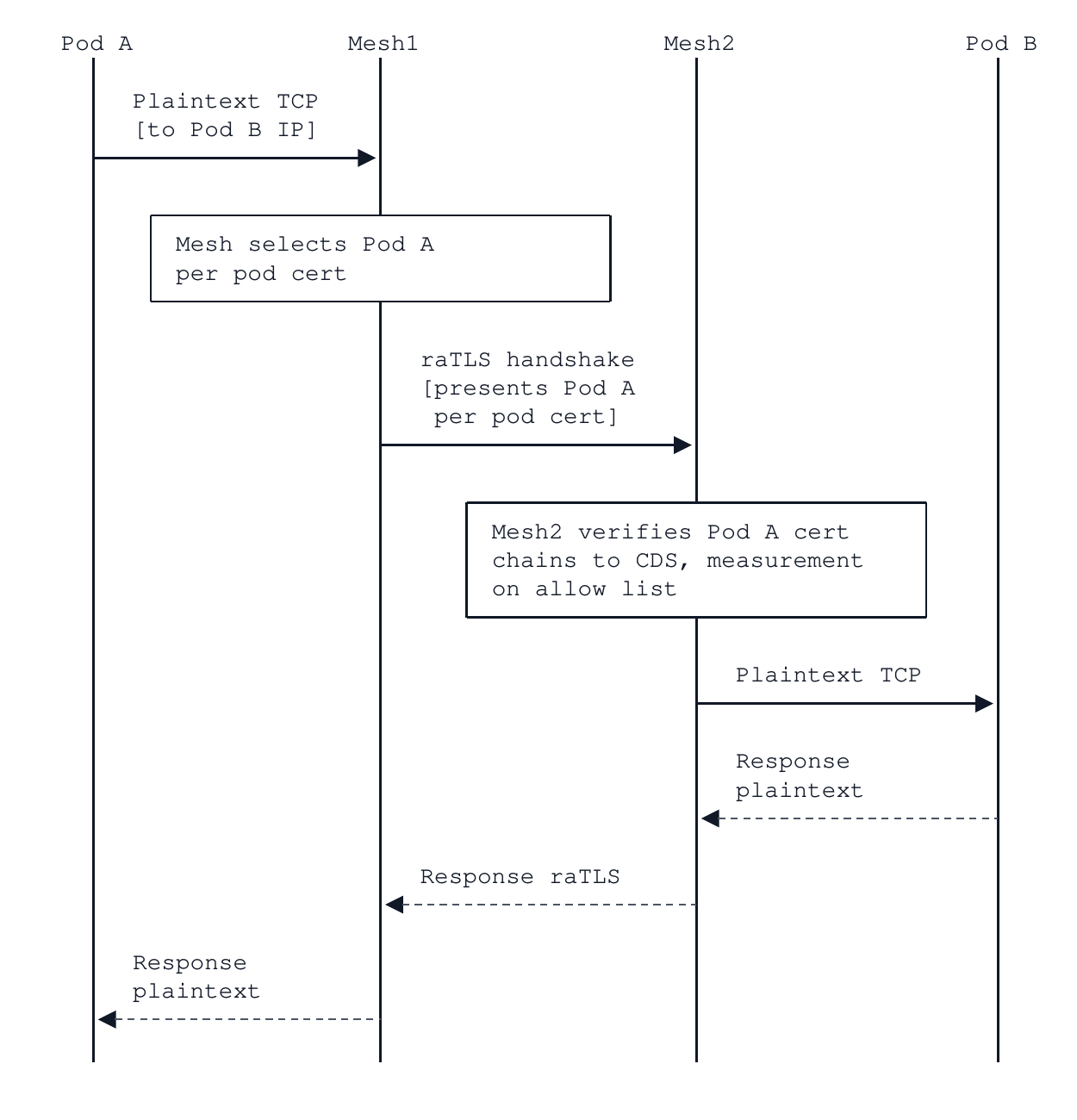}
\caption{Pod-to-pod mesh flow. Mesh1 and Mesh2 are the node-level mesh proxies on Pod A's and Pod B's nodes respectively. The workload observes plaintext on both ends; raTLS with per-pod certificates is terminated at the node-level proxy on each side.}
\label{fig:mesh-flow}
\end{figure}

\subsubsection{Identity Granularity}

Every pod receives its own CDS-issued certificate (the pod is always the unit of identity), but there are two reasonable choices for where that certificate actually lives on the wire, and they differ in how much machinery the customer has to tolerate in their pod spec.

\textbf{Per-pod via node-level proxy.} A single mesh proxy runs on each node (typically as a DaemonSet) and holds the certificates for every pod on that node. When a pod opens a connection, the proxy selects the pod's certificate and presents it to the peer. The receiving side sees pod-level identity; the customer sees no additional sidecars in their workload. The trade-off is concentrated in the proxy, which now manages a growing and churning set of certificates rather than a single identity.

\textbf{Per-pod via sidecar.} Each pod gets its own mesh proxy colocated as a sidecar container. Each sidecar holds exactly one certificate. This is the cleanest arrangement from an identity-accounting perspective and mirrors the Istio deployment model, but it requires injecting a sidecar into every customer pod. That is costly for deployments that were not designed around service-mesh conventions, and at odds with C8s's preference for minimal workload-side disruption.

The on-the-wire semantics are identical in both cases. A raTLS handshake presents a per-pod certificate that chains to the CDS CA and corresponds to a measurement on the allow-list. The choice is operational. Deployments with very high pod churn, or those already standardized on per-pod sidecars for other reasons, may prefer the sidecar arrangement; most prefer the shared proxy.

\begin{figure}[H]
\centering
\makebox[\linewidth]{\includegraphics[width=1.1\linewidth]{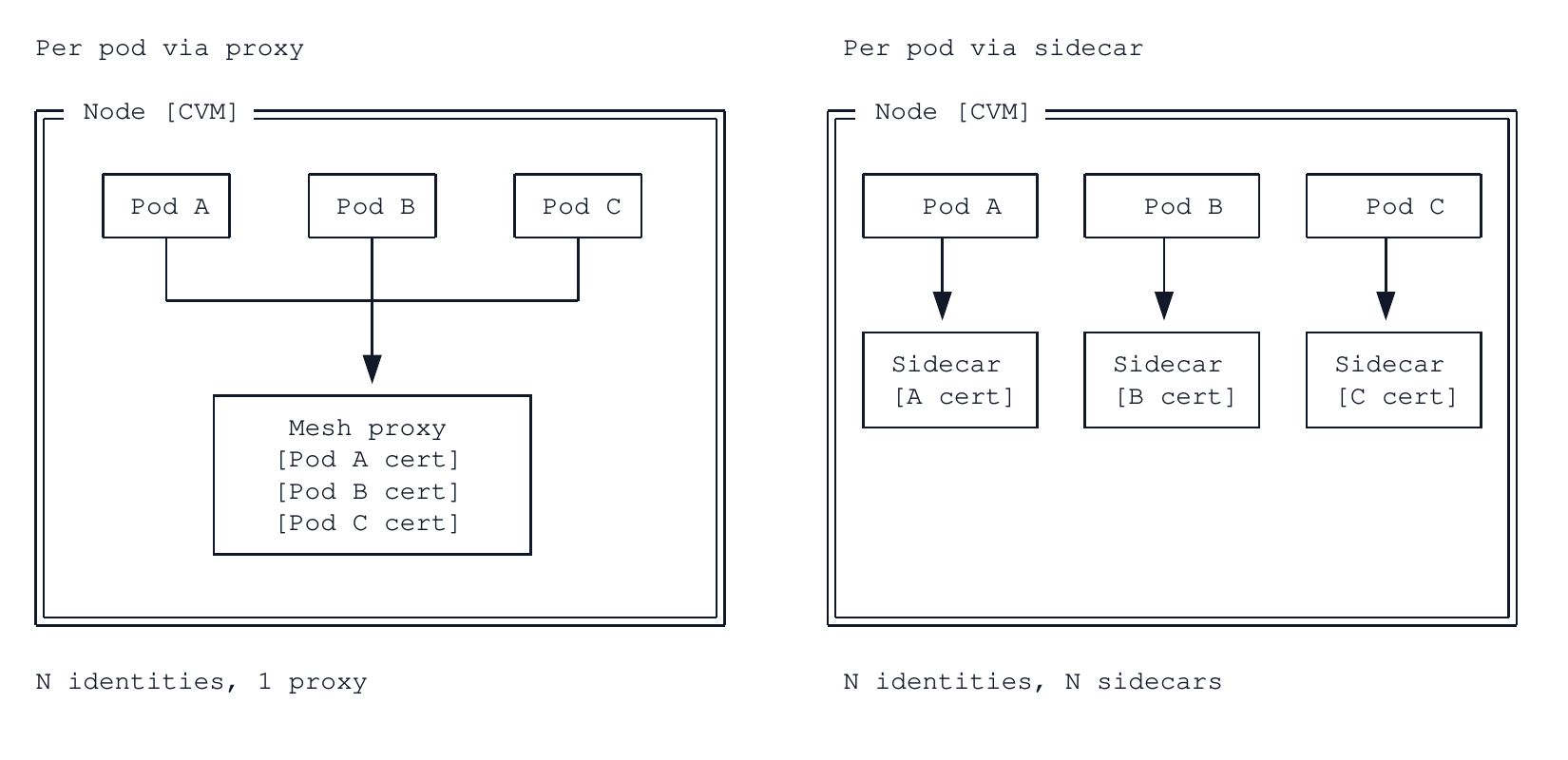}}
\caption{Two identity-holding arrangements. Both preserve pod-level identity on the wire; they differ in whether the certificate lives in a shared node-level proxy or in a sidecar colocated with each pod.}
\label{fig:identity-granularity}
\end{figure}

\subsubsection{Freshness and Binding}
\label{sec:freshness}

Two concerns recur in attested TLS designs. Both have seen formal analysis and both admit a straightforward architectural answer. C8s addresses them by construction rather than as after-the-fact hardening.

\textbf{Evidence-to-channel binding.} An attestation report proves that a particular code measurement ran at some point on genuine hardware. It does not, on its own, prove that the same hardware is one end of a given TLS session. This is the \emph{channel binding} problem in the sense of RFC 5056~\cite{ref:channel-binding}, applied to attestation evidence. If evidence were accepted independently of the session it arrived on, a compromised pod with a valid report, or an adversary who captured a report in transit, could present that report while proxying a session to a different key. The bind is what makes the report mean ``this connection'' rather than ``some connection somewhere.''

C8s binds evidence to the session at certificate issuance time. The attested pod generates an ephemeral key pair inside its CVM's hardware-encrypted memory, and the attestation report it presents to the CDS includes a digest over that public key in a dedicated report field. The CDS verifies the digest before issuing the certificate and embeds the same public key in the certificate it returns. A peer completing a raTLS handshake against this certificate is therefore talking to the holder of the private key that was inside the TEE at attestation time. The report cannot be reused with a different key because the digest in the report field would not match.

This puts the binding in the attestation evidence itself, not in a TLS extension or a separate channel. The CDS refuses to issue a certificate if the report's key digest does not match the presented public key, so a pod that failed to bind correctly never gets a certificate in the first place. Peers do not need to re-verify the binding on every connection; they verify the CDS's signature on the certificate, and the binding property is implied by the CDS's issuance policy.

\textbf{Certificate freshness.} A bound certificate still represents attestation at a point in the past. If a pod's certificate is valid for days or weeks, the gap between ``measurement verified'' and ``connection accepted'' grows accordingly, and a post-attestation compromise (key extraction via a side channel, a firmware-level flaw disclosed after issuance, an operator exfiltrating a running pod's key) has longer to do damage.

C8s issues short-lived certificates as the default posture. Lifetimes are measured in hours, not days, so the window in which a compromised or stale certificate is accepted is bounded. Short lifetimes are cheap in this architecture because issuance is automated and attested pods re-attest on renewal. The CDS re-verifies the pod's current attestation report, re-checks the key binding, and re-checks the measurement against the allow-list before issuing a replacement. If the pod's TCB has fallen below the minimum (because a firmware advisory has raised the floor), or the measurement is no longer on the allow-list (because the image has been revoked), renewal fails and the pod drops out of the mesh at the old certificate's expiry. There is no separate revocation channel to operate or to fail. Expiry plus re-attestation is the revocation mechanism.

The net effect is that C8s's raTLS identity is bound to the session by construction (via key-digest-in-report) and short-lived by default (to bound staleness). Expiry plus re-attestation provides the freshness guarantee for the mesh; raTLS deliberately does not reuse the externally facing freshness beacon (\S\ref{sec:default-tls}) to avoid confused-deputy issues from sharing a single signing operation across two protocols.

\subsection{Client Connection Protocol}

The client-side protocol decouples attestation verification from the transport layer. Rather than relying on TLS extensions (such as Exported Keying Material) to bind attestation to the transport session, the architecture moves attestation verification to the application layer. This decoupling improves client compatibility, so the attestation-aware client works in any HTTP client, including browsers, without requiring TLS extension support.

\subsubsection{Bootstrapping}
\label{sec:client-bootstrap}

On first connection, the client retrieves two artifacts from the CDS. The first is the CDS's own attestation report. The second is the CDS-signed manifest, which is the same allow-list the CDS uses internally for image policy enforcement (\S\ref{sec:cds}, \S\ref{sec:nri}), exposed for client-side use.

The CDS attestation report anchors the trust chain. The client verifies it against the hardware vendor's root of trust, checks that the measurement matches the expected CDS image, and confirms that the public key bound by the report is the same key the manifest is signed under. From that point the client trusts the CDS public key on the basis of its own attestation check, not on operator assertion. The same CDS public key serves a second purpose. It is the trust anchor for any CDS-issued certificate the client subsequently encounters, including the certificate presented by whatever cluster component the client establishes a direct connection to (the ingress router in the default deployment, \S\ref{sec:ingress}).

Pool members are trusted transitively through the CDS signature on the manifest and the raTLS chain inside the cluster. The component the client connects to directly is attested separately on each connection window via the freshness beacon (\S\ref{sec:default-tls}).

The manifest contains:

\begin{itemize}
\item The CDS public key, bound by the CDS attestation the client has just verified. It is used both to verify the manifest signature and to anchor any CDS-issued certificate the client subsequently sees.
\item The allow-list of permitted measurements, including the expected measurement for the component the client connects to directly and any minimum TCB versions or other appraisal policy fields the CDS enforces.
\end{itemize}

Because the client has already attested the CDS, verifying the manifest signature is sufficient to trust its contents. The client uses the allow-list to recognize a valid measurement when one is presented in an attestation report, but performs no per-machine appraisal.

Verified data is cached locally. Subsequent requests use the cached values, with periodic refresh to keep up with allow-list changes. The bootstrapping cost is amortized, so only the first connection incurs the full CDS attestation.

\subsubsection{Default: Traditional TLS + Attestation}
\label{sec:default-tls}

In the default configuration, the client establishes a connection to the cluster's ingress router over ordinary HTTPS. The ingress presents (a) a public-CA-issued TLS certificate whose key was generated and is held inside the ingress's TEE, and (b) an attestation report. The client verifies that the TLS certificate's public key matches the bound public key in the attestation report, that the attestation's measurement matches the ingress identity in the CDS-signed manifest, and that the attestation is fresh under the freshness beacon described below. Once these checks pass, the client treats the open TLS session as bound to an attested ingress and reuses it for subsequent requests until the freshness window or session lifetime elapses. A malicious control plane could in principle forge a public-CA TLS certificate, but it could not produce a valid attestation binding the certificate's key to an attested ingress measurement, so a verifying client refuses to use the session.

The chain of trust runs end to end:

\begin{enumerate}
\item The operator attests the CDS on cluster install (\S\ref{sec:cds-bootstrap}).
\item The client attests the CDS on first connection (\S\ref{sec:client-bootstrap}).
\item The CDS countersigns the freshness beacon used by the ingress.
\item The ingress's attestation binds its TLS certificate's public key to the measured ingress.
\item The ingress opens a raTLS connection to the destination pod using a CDS-issued mesh certificate, which the CDS issued only after the pod's own attestation.
\end{enumerate}

Every hop after the client's TLS termination is therefore between attested peers whose identities chain back to the same CDS the client itself has attested.

\textbf{Freshness beacon.} TEE attestation signing has limited throughput, so requiring a fresh attestation per request is infeasible at scale. The CDS produces a freshness beacon (a timestamp-based freshness mechanism in the sense of RFC 9334 \S10~\cite{ref:rats}, with the CDS signature serving as the unforgeable nonce) on demand that any Verifier can use to bound the staleness of a presented attestation:

\begin{enumerate}
\item When an attested workload (typically the ingress) detects that its current attestation is approaching the freshness window's edge, it requests a fresh beacon from the CDS over its existing raTLS connection. The CDS signs the current timestamp $T$ with its signing key, applying a fixed domain separator (e.g., \texttt{"c8s/freshness-beacon/v1"} $\|$ $T$) so the beacon-signing operation cannot be confused with any other CDS signature. It returns $\mathrm{sig}_\mathrm{CDS}(T)$ to the caller.
\item The workload asks the TEE to produce a new attestation whose \texttt{REPORT\_DATA} field carries $\mathrm{sig}_\mathrm{CDS}(T)$ and whose bound-pubkey field carries the workload's TLS public key.
\item When a client connects, the workload presents the attestation report along with $(T, \mathrm{sig}_\mathrm{CDS}(T))$. The client then checks each of the following:
\begin{itemize}
\item $\mathrm{sig}_\mathrm{CDS}(T)$ is a valid CDS signature over the domain-separated input under the CDS public key established at trust bootstrap.
\item $T$ falls within the configured freshness window.
\item The attestation's \texttt{REPORT\_DATA} equals $\mathrm{sig}_\mathrm{CDS}(T)$ exactly.
\item The attestation's bound public key matches the TLS certificate's public key.
\item The measurement is on the allow-list and the TCB meets the minimum.
\end{itemize}
\end{enumerate}

A CDS signature is unforgeable to any party that does not hold the CDS private key, so it functions as an unforgeable nonce. Combined with the timestamp, the nonce bounds when the attestation could have been generated. The verifier needs no out-of-band data, since the timestamp and signature both arrive in-band with the attestation, and the CDS public key was already cached during bootstrapping (\S\ref{sec:client-bootstrap}). The window is operator-configurable and typically on the order of minutes, large enough to amortize the cost of TEE attestation generation given hardware throughput limits, and small enough to limit the impact of a post-attestation compromise. If a workload's attestation goes too stale, clients refuse to talk to it, mirroring the behavior of expired raTLS certificates inside the mesh.

\subsubsection{Optional: Multi-Recipient Client-Side Encryption}
\label{sec:client-mre}

For clients that want end-to-end encryption to the destination pod's own key, typically because they don't trust the ingress router to terminate TLS, or want crypto-agility for the payload independent of the transport, C8s supports client-side multi-recipient encryption (\S\ref{sec:mre}). The client encrypts the payload to the public keys of attested pods in the target pool, using a hybrid scheme. The body is encrypted once with a fresh symmetric key, and that key is wrapped to each recipient's public key. The wrapped header carries, per recipient:

\begin{itemize}
\item The public-key material the corresponding pod uses to derive the symmetric decryption key, and
\item A plaintext routing hint (e.g., the pod's hostname), used by the Encrypted Ingress Router (\S\ref{sec:ingress}) for forwarding decisions.
\end{itemize}

Overhead per additional recipient is approximately 64 bytes in the header. Clients may encrypt to a subset of the pool rather than its entirety, trading routing flexibility for smaller headers. The reference implementation uses AGE (\S\ref{sec:mre}), but the protocol is not bound to a specific format. Threshold, hybrid, or post-quantum schemes can be substituted without changes elsewhere. This path is available to clients that opt in, and it requires the Encrypted Ingress Router variant on the cluster side.

\begin{figure}[H]
\centering
\includegraphics[width=0.88\linewidth]{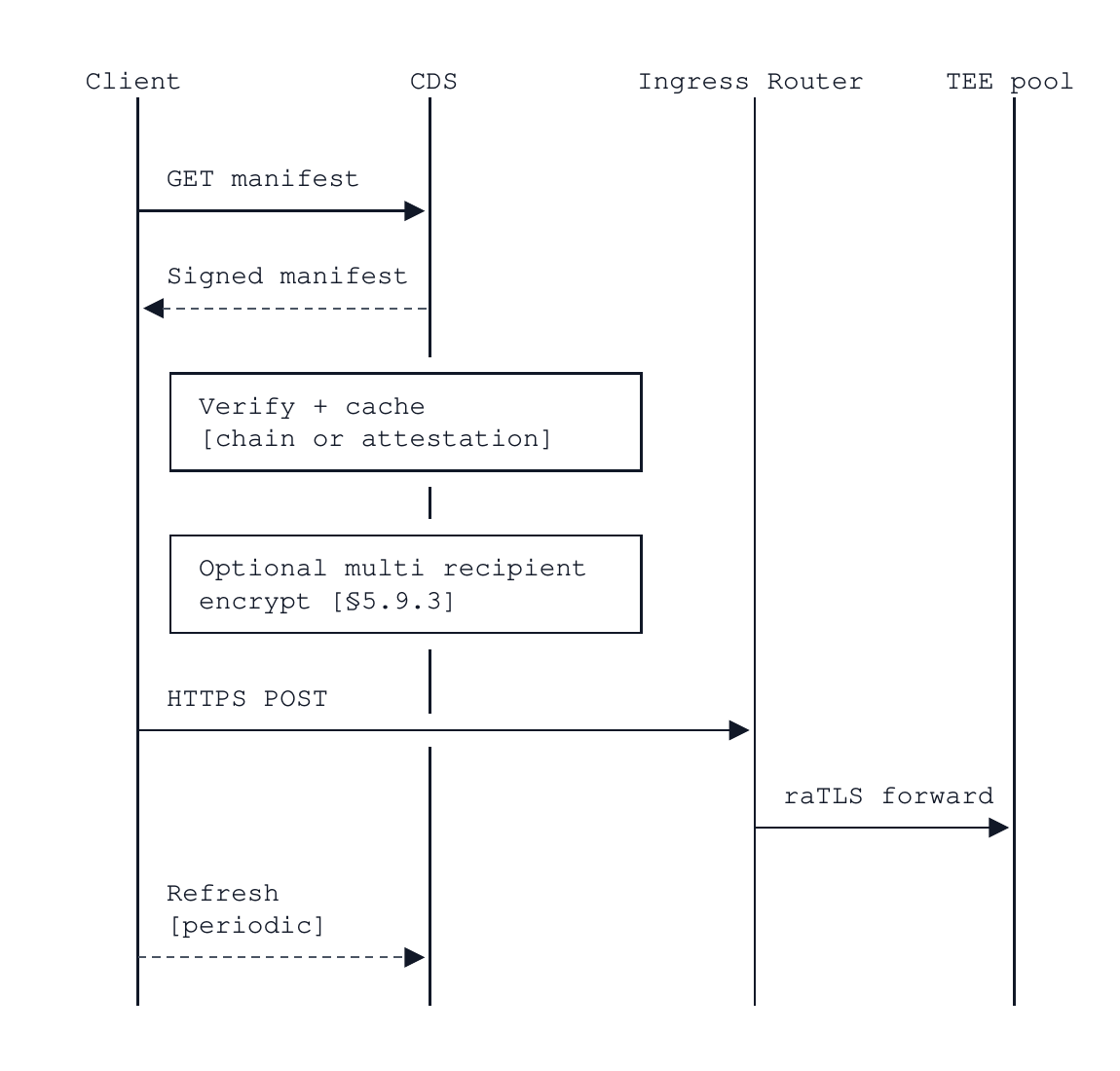}
\caption{Client connection flow. The manifest fetch is amortized through caching and periodic refresh, and each request only exercises the lower half of the diagram. The multi-recipient encryption step is optional. By default the client submits the request over ordinary HTTPS and the ingress router does the raTLS forwarding.}
\label{fig:client-bootstrap}
\end{figure}

\subsubsection{Response Path}

Attested workloads include attestation metadata in response headers, enabling the client to verify that the response originated from within the trust boundary. The response body is encrypted to the client using keys established during the session. The full request-response cycle operates over standard HTTPS; no custom transport protocol is required.

\subsection{Ingress Router}
\label{sec:ingress}

External clients need to reach a specific pod inside the trust boundary. C8s offers two interchangeable ways to do this, both running inside the cluster's trust boundary. They differ in how much they can see of the request payload.

The router plays the same data-plane role as a standard Kubernetes ingress controller (nginx, Envoy, HAProxy, or a cloud-native equivalent). It accepts external traffic, terminates a public TLS endpoint, and forwards to a backing pool. Two things distinguish it from those primitives. First, the entire data path runs inside a CVM, so traffic in flight is never visible to the host or the control plane. Second, it refuses to forward to any peer that does not present a valid CDS-issued raTLS certificate, so a control-plane edit that points a Kubernetes Service at an unattested endpoint cannot redirect traffic.

\subsubsection{Default: raTLS pass-through}
\label{sec:default-ingress}

The default ingress router is an attested load balancer running inside a CVM alongside the rest of the C8s control-boundary components. Externally, it presents a public-CA-issued TLS certificate whose key is generated and held inside the ingress's TEE, paired with a recent attestation that the client verifies (\S\ref{sec:default-tls}). The TLS certificate itself is obtained through ordinary mechanisms (e.g., ACME running inside the attested ingress to bind the cert to the TEE-internal key). The precise issuance workflow is out of scope here. Internally, the ingress holds a CDS-issued raTLS mesh certificate for talking to other attested pods, and for each incoming request it opens a raTLS connection to the target pod over the mesh.

\textbf{Service discovery and downstream membership.} Pool membership comes from the standard Kubernetes Service object the router watches; the router learns about new endpoints, scaling events, and pod churn through the same kube-apiserver mechanisms an ordinary ingress controller uses. A control plane that adds a malicious endpoint cannot weaponize this, because the raTLS handshake to that endpoint will fail unless the endpoint holds a CDS-issued mesh certificate, and the CDS only issues to attested measurements on the allow-list. Health-checking is similarly TLS-aware: a peer that fails the raTLS handshake (revoked, expired, or unattested) is dropped from the live pool until it can present a valid certificate again. Failover across multiple healthy peers uses the same load-balancing strategies any modern ingress supports; the difference is the admission criterion, not the load-balancing algorithm.

The chain of trust runs end to end. The operator attests the CDS at bootstrap (\S\ref{sec:cds-bootstrap}). The CDS in turn countersigns the ingress's freshness beacon and validates its attestation against the allow-list. The ingress then validates the destination pod's mesh certificate. From the client's perspective the router is a normal HTTPS endpoint that additionally presents an attestation. From the cluster's perspective every hop after TLS termination is raTLS between attested peers.

This gives end-to-end semantics without requiring the client to encrypt payloads to specific pod keys. The client still runs the standard raTLS verification path (attestation signature, freshness beacon, measurement allow-list, TLS key binding), but it does not need to manage per-pod recipient keys or wrap payloads itself the way it does in the multi-recipient encryption variant (\S\ref{sec:client-mre}). The request is in plaintext only inside the router's CVM memory (briefly, while it is being forwarded) and inside the destination pod's CVM memory. Neither endpoint is reachable from the hypervisor, host OS, or the control plane. Because the router itself is attested and mesh-resident, compromising the cluster's ordinary Kubernetes ingress does not yield access to request bodies. The attacker would have to compromise the router's CVM, which is the same threat model as compromising any other attested workload.

\begin{figure}[H]
\centering
\includegraphics[width=0.78\linewidth]{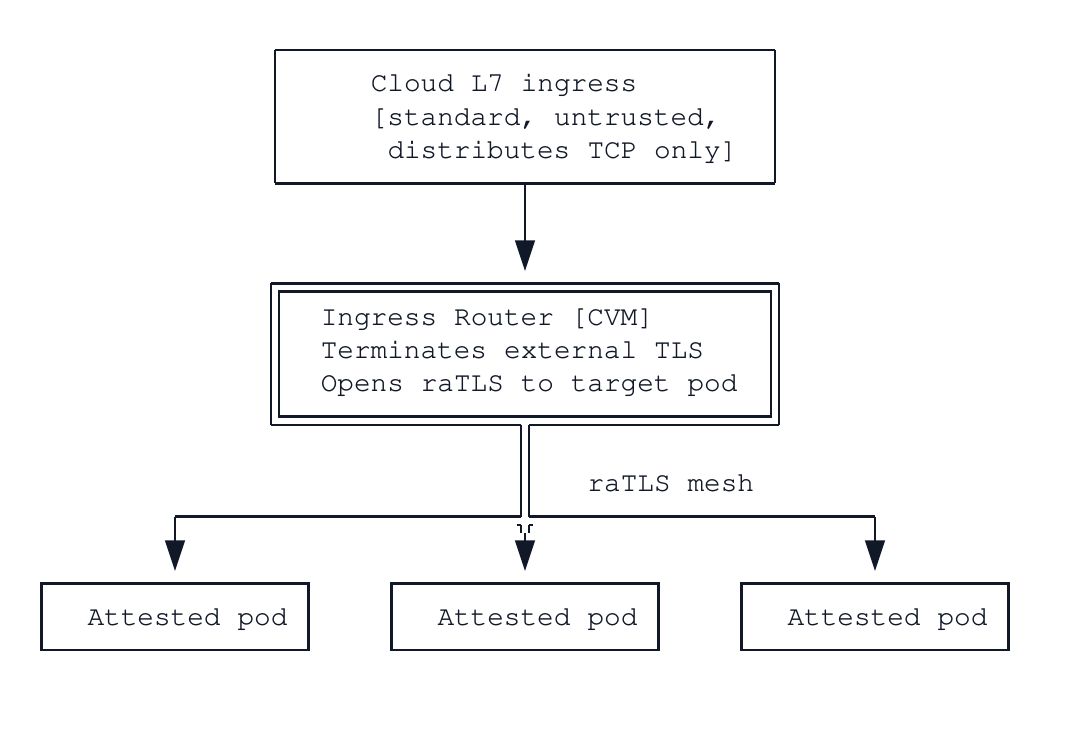}
\caption{Default ingress path. The ingress router is itself an attested CVM; it terminates external TLS, then relays requests to the destination pod over raTLS. No other cluster component ever sees request payloads in plaintext.}
\label{fig:ingress-default}
\end{figure}

\subsubsection{Optional: Encrypted Ingress Router}
\label{sec:encrypted-ingress}

Some deployments want a stronger property, where the ingress never sees plaintext, not even briefly inside its own CVM. This is useful when the cluster operator is not the same entity as the workload owner (the workload owner trusts their own attested pod but not the router's), when multi-recipient crypto agility is valuable (threshold, hybrid, post-quantum), or as defense in depth against a router compromise.

For these cases C8s ships an \textbf{Encrypted Ingress Router} variant, implemented as a per-node DaemonSet. Clients encrypt requests with multi-recipient encryption (\S\ref{sec:mre}, \S\ref{sec:client-mre}) to a pool of attested pods, submit the ciphertext through ordinary Kubernetes ingress, and the router parses the per-recipient header, extracts the plaintext routing hints from the recipient stanzas, and forwards the still-encrypted payload to the destination pod. The router never decrypts the payload body; it operates exclusively on the routing metadata in the header. A compromise of the router exposes routing metadata only, and workload data remains encrypted end-to-end to the destination pod.

On each node the router runs as an attested pod. It boots with a measured image, attests to the CDS, receives a per-pod certificate, and participates in the raTLS mesh. When a request arrives:

\begin{enumerate}
\item The router parses the encrypted payload header and extracts the plaintext routing hints from the recipient stanzas.
\item If any hint matches an attested pod on the local node, the router forwards the still-encrypted payload to that pod.
\item Otherwise, the router forwards the payload across the raTLS mesh to a peer Encrypted Ingress Router on the node that hosts a matching pod.
\item The destination attested pod decrypts the payload with its private key (held only in CVM-encrypted memory) and processes the request.
\end{enumerate}

\begin{figure}[H]
\centering
\includegraphics[width=0.98\linewidth]{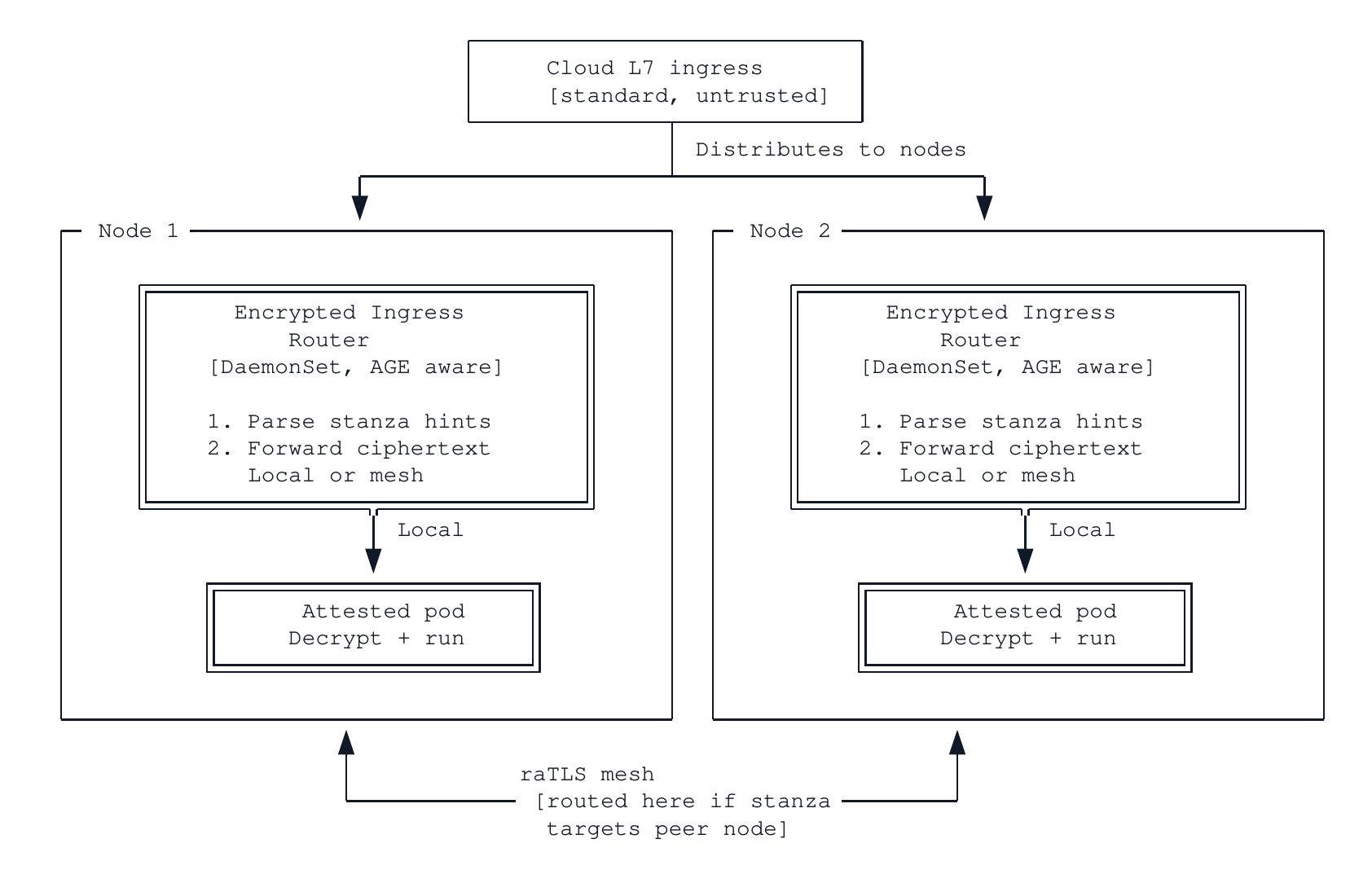}
\caption{Encrypted Ingress Router topology. Cloud ingress distributes TCP to nodes without understanding the encryption format. A per-node router (attested DaemonSet) parses recipient hints and forwards ciphertext to a local pod or a peer router on another node. No decryption happens in the router; at most one intra-cluster hop across the mesh.}
\label{fig:ingress-encrypted}
\end{figure}

\textbf{Why per-node rather than centralized.} A single centralized payload-aware load balancer would work, but it concentrates routing in one component and adds a hop that duplicates what Kubernetes ingress already does. Running the Encrypted Ingress Router as a per-node DaemonSet scales linearly with cluster size, eliminates the single-component concentration, and lets the cloud LB play its ordinary role (distributing TCP to nodes) without C8s-specific behavior.

\textbf{Why not a sidecar in each customer pod.} Sidecars would require injecting C8s machinery into every customer workload manifest, which cuts against the architecture's preference for leaving customer workloads unchanged. The DaemonSet model lives at the node level and is invisible to the customer's pod spec.

\textbf{Response path.} Responses are encrypted by the destination pod to the client's public key (carried in the request) and travel back through the same path, from pod to local router to cloud ingress to client. The mesh and the router both see ciphertext on the return leg as well.

\subsubsection{Choosing between the two}

The two variants are compatible with the rest of the architecture; the choice is deployment-time and can be made per cluster or per workload. The default raTLS pass-through is simpler for operators and requires no client-side cryptography. The Encrypted Ingress Router is appropriate when the threat model includes an untrusted or potentially-compromised ingress component, or when the workload's clients already speak a multi-recipient encryption format for other reasons. Both variants can coexist in the same cluster.

\subsection{Attested Build System (Kettle)}

Kettle is the build pipeline that produces container images with attestable digests. Builds execute inside a TEE, and the build environment is itself attested, so the output of a Kettle build is a container image paired with a hardware-signed attestation report binding the build inputs to the output digest. This extends the chain of trust from source code to running container. Without an attested build system, an attacker who compromises the CI/CD pipeline could produce a malicious image and add it to the policy manifest.

The policy manifest consumed by image policy enforcement is signed by a key that only Kettle possesses within its own TEE. The CDS trusts this manifest because it can verify the signature against Kettle's attested public key. An untrusted operator cannot substitute a different manifest without access to Kettle's signing key, which exists only within Kettle's hardware-encrypted memory. The operator can deploy the manifest, but they cannot forge it. A detailed treatment of Kettle's design and implementation is the subject of a separate publication.

\section{Request Lifecycle}

This section traces a single request through the system, identifying where each security guarantee applies.

\subsection{Preconditions}

Before the first request, the following setup has occurred:

\begin{enumerate}
\item The CDS has been deployed inside a CVM and attested by the operator.
\item The CVM substrate that will host the workload has booted. In the default pod-level configuration, the C8s pod runtime is installed on each host and the measured per-pod configuration has been published. In a node-level configuration, worker nodes have booted as CVMs and each node's secure processor has computed a launch measurement. Either way, the CVM that will host the workload has a pre-computable launch digest that the CDS's allow-list recognizes.
\item The CVM substrate has attested to the CDS. In the pod-level configuration this happens once per pod at start, directly; in the node-level configuration the node attests first and then composes per-pod evidence on top.
\item Image-gap gating is in place. In the pod-level configuration, measured image pinning, per-customer workload signing, or both are baked into the pod's measured configuration and enforced by the in-pod agent at pull time. In the node-level configuration, the pod attestation agent, NRI image policy enforcer, and raTLS mesh interceptors run as DaemonSets, have attested to the CDS as pods, and hold their per-pod mesh certificates and the signed policy manifest.
\item Workload pods have been scheduled and started. Each workload pod has presented attestation evidence to the CDS (a direct per-pod attestation in the default configuration, or composed node-and-workload evidence in the node-level configuration) and received its own per-pod mesh certificate. All inter-pod connections use these certificates.
\item Any application secrets (if encrypted) have been decrypted inside TEE memory using keys released by the CDS to the attested pod under the brokering flow described in \S\ref{sec:key-broker}. The key exists only in the pod's hardware-encrypted memory; it is not persisted to pod storage.
\item The ingress router holds a public-CA-issued TLS certificate bound to its TEE-internal key, plus a recent attestation that incorporates the latest CDS freshness beacon. It also holds a CDS-issued raTLS mesh certificate for internal traffic.
\end{enumerate}

The following diagram traces a single request through the pipeline. Each dashed band corresponds to one of the subsections below; readers can use it as a map for the narrative that follows.

\begin{figure}[H]
\centering
\makebox[\linewidth]{\includegraphics[width=1.2\linewidth]{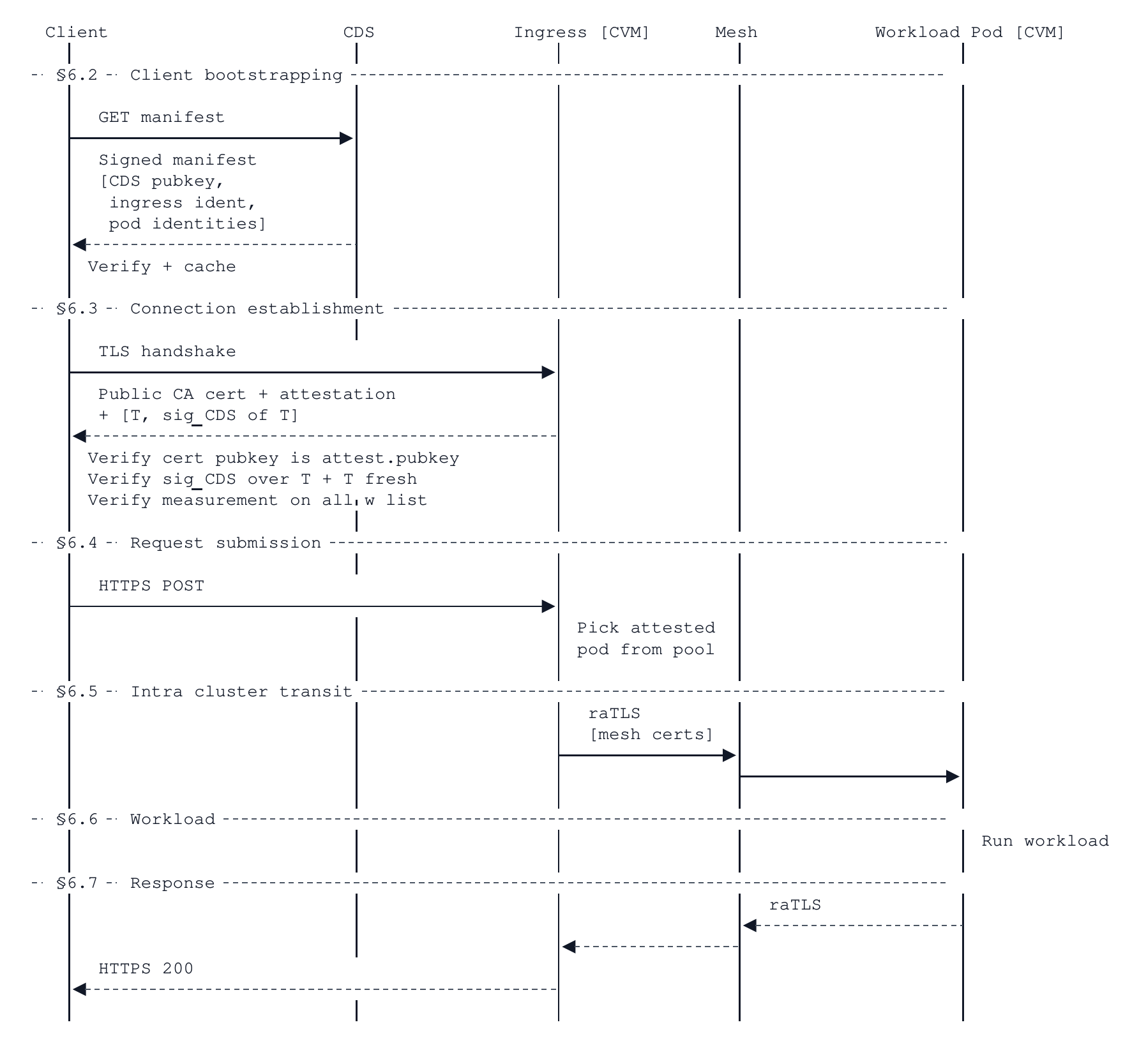}}
\caption{End-to-end request lifecycle in the default configuration. Dashed bands mark the subsection boundaries used in the rest of this section. The optional client-side multi-recipient encryption path (\S\ref{sec:client-mre}) replaces the \texttt{HTTPS POST} step with a ciphertext payload routed by hostname hints (see \S\ref{sec:encrypted-ingress}).}
\label{fig:request-lifecycle}
\end{figure}

\subsection{Client Bootstrapping}

The client's attestation SDK contacts the CDS and retrieves the CDS-signed manifest containing the CDS public key, the ingress identity (its expected measurement and TLS public key), and the public keys and attestation references for the attested pod pool. The SDK verifies the manifest (either via certificate chain validation against the CDS public key or by independent attestation verification) and caches the verified data. This step occurs once and is amortized across subsequent requests; periodic refresh covers pool membership changes.

\subsection{Connection Establishment}

The client opens a TLS connection to the ingress router. The ingress presents (a) a public-CA-issued TLS certificate whose key was generated and is held inside the ingress's TEE, and (b) a recent attestation report along with $(T, \mathrm{sig}_\mathrm{CDS}(T))$ from the latest freshness beacon. The client verifies that the TLS certificate's public key matches the bound public key in the attestation, that the attestation's measurement matches the ingress identity in the cached manifest, and that $\mathrm{sig}_\mathrm{CDS}(T)$ is a valid CDS signature over a sufficiently recent $T$ (\S\ref{sec:default-tls}). Once these checks pass, the open TLS session is treated as bound to an attested ingress and reused for subsequent requests until the freshness window or session lifetime elapses.

\subsection{Request Submission}

The client sends the request body over the established TLS connection. Inside the ingress's CVM-encrypted memory, the request is briefly in plaintext while the ingress selects an attested pod from the target pool. (Clients that prefer to keep the request body opaque to the ingress can use the optional client-side multi-recipient encryption path described in \S\ref{sec:client-mre} with the Encrypted Ingress Router variant in \S\ref{sec:encrypted-ingress}; the ingress then operates only on the encrypted payload's routing hints.)

\subsection{Intra-Cluster Transit}

The ingress opens a raTLS connection to the destination pod. Both endpoints hold CDS-issued per-pod mesh certificates whose key digests bound them to attestations the CDS verified. The mesh encrypts the request in transit; the underlying network sees only TLS ciphertext between attested peers.

\subsection{Workload Execution}

The destination pod receives the request inside its CVM-encrypted memory and processes it as it would on any standard deployment. Inputs, intermediate state, working buffers, and any sensitive artifacts the workload has loaded (model weights, decryption keys, dataset shards, downstream credentials) reside in memory encrypted with keys the hypervisor never possesses. The execution itself is whatever the workload is --- a database query, a batch job, an LLM generation step (tokenization, attention, KV cache updates, token sampling), a signing operation, etc. (Where the optional client-side multi-recipient encryption path is in use, the pod additionally decrypts the per-recipient envelope using its private key, held only in CVM-encrypted memory, before processing the request.)

\subsection{Response}

The workload response is returned through the same path. It travels from pod to ingress via the raTLS mesh, then from ingress to client over the established TLS session. Attestation metadata is included in response headers, enabling the client to confirm the response originated from an attested TEE within the verified pool.

\section{Untrusted Control Plane Compatibility}

\subsection{Design Rationale}

Managed Kubernetes services (AKS, EKS, GKE) operate the control plane as a service. The operator does not control the control plane's code, cannot run it in a TEE, and cannot attest it. Requiring the control plane to be inside the trust boundary would exclude the majority of production Kubernetes deployments.

The architecture deliberately places the control plane outside the trust boundary. The control plane retains its scheduling and orchestration functions but is stripped of any role in confidentiality enforcement. Trust anchors are moved to the node: the NRI image policy enforcer gates workload execution, the CDS gates identity issuance, and the raTLS mesh gates network participation.

\subsection{Blast Radius of Control Plane Compromise}

A compromised or malicious control plane is limited to denial of service. It can refuse to schedule workloads, terminate pods, or disrupt service discovery. It cannot:

\begin{itemize}
\item Access workload data (encrypted in TEE memory and in transit).
\item Forge attestation reports (requires hardware-rooted signing keys).
\item Inject unauthorized workloads (rejected by the NRI enforcer).
\item Obtain mesh credentials (issued only by the CDS to attested pods).
\item Access model decryption keys (released by the CDS only to attested pods; present only in the pod's hardware-encrypted memory, never persisted or exposed to the control plane).
\end{itemize}

When the trust boundary is drawn at the pod rather than the node, the ``control plane outside the trust boundary'' property extends even to managed Kubernetes clusters whose nodes have no TEE support at all. The pod's CVM is provisioned per-pod (for example as a confidential cloud VM), so the cluster's node image and scheduler stay on the untrusted side without weakening any of the guarantees above.

\subsection{Self-Hosted Control Plane}

Organizations that operate their own control plane may optionally deploy it inside CVMs and attest it alongside worker nodes. This brings the control plane inside the trust boundary and protects cluster state (scheduling metadata, etcd contents, configuration) from the infrastructure operator. This option is not available on managed Kubernetes services and adds the operational cost of running the control-plane components (API server, scheduler, etcd, controller manager) as attested CVMs, with the corresponding attestation, upgrade, and key-management workflows applied to each. It is not required for the confidentiality guarantees described in this paper.

\section{Configurable Boundaries}

The architecture is not a fixed blueprint. Components can be combined, extended, or omitted depending on the threat model and operational requirements.

\subsection{Minimal Deployment}

The recommended starting point includes:

\begin{itemize}
\item A CVM substrate attested to the CDS, with per-pod CVMs provisioned by the C8s pod runtime by default (\S\ref{sec:pod-cvm}), or CVM worker nodes where the boundary is drawn at the node (\S\ref{sec:node-cvm}).
\item Measurement-bound workload gating, implemented as measured image pinning or per-customer workload signing inside each pod's configuration in the default pod-level setup, or as NRI image policy enforcement on each node (via DaemonSet) in a node-level setup; both can coexist in a mixed cluster.
\item raTLS mesh with CDS-issued per-pod certificates for all inter-pod traffic.
\item An attested ingress with a public-CA-issued TLS certificate bound to a TEE-internal key, and the CDS freshness beacon (\S\ref{sec:default-tls}) for client-side staleness checks.
\item Control plane remains untrusted.
\end{itemize}

This configuration is compatible with managed Kubernetes and provides the full set of confidentiality guarantees described in this paper.

\subsection{Oblivious HTTP (OHTTP)}

Standard TLS protects request contents but exposes metadata such as which clients connect, when, and how often. For applications where traffic analysis is a privacy concern, such as healthcare or legal queries, the architecture supports an OHTTP~\cite{ref:ohttp} relay layer. The relay observes client IP addresses but cannot read request payloads (which are encrypted to the attested gateway under multi-recipient encryption, \S\ref{sec:client-mre}). The gateway processes requests but sees only the relay's address, not the originating client. No single party learns both identity and content.

\begin{figure}[H]
\centering
\includegraphics[width=\linewidth]{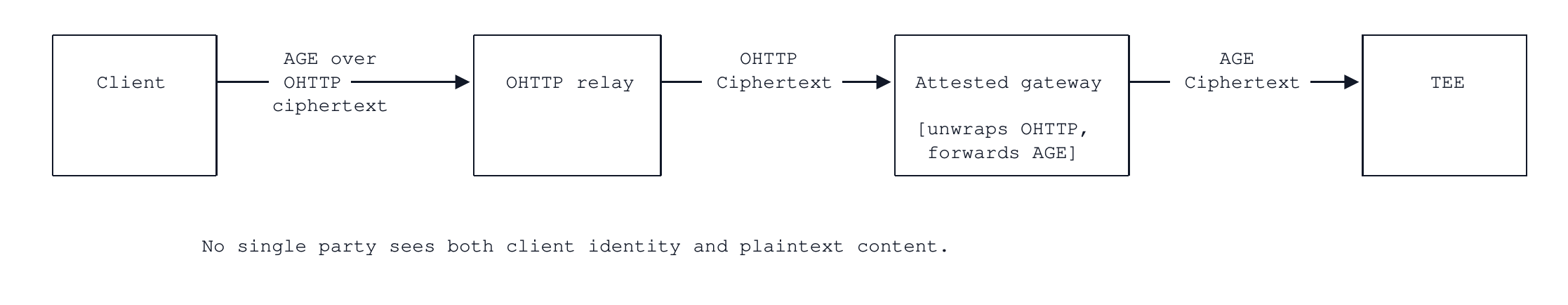}
\caption{OHTTP relay topology. Each hop sees either the client identity or the payload, never both; the relay cannot decrypt the OHTTP envelope, and the gateway cannot see the originating client.}
\label{fig:ohttp}
\end{figure}

\section{Discussion}

\subsection{Trust Assumptions}

The architecture concentrates trust in hardware manufacturers (AMD, Intel) and the code running inside measured TEEs. The CDS is explicitly trusted as the root of the certificate chain.

A CDS compromise would allow an adversary to issue fraudulent certificates and, through them, impersonate attested workloads to peers that accept CDS-issued identity. It would not retroactively decrypt past traffic, break in-flight sessions protected by ephemeral keys, or reach CVM memory whose keys the CDS has never held, because the CDS is deliberately scoped to CA signing material and attestation-gated brokering (see key-minimization discussion in \S\ref{sec:cds}).

This trust is further mitigated by the CDS running inside a CVM whose attestation the operator manually verifies at deployment time.

If the hardware manufacturer is compromised or malicious, the attestation guarantees do not hold. This is a narrower trust surface than the standard model (which additionally trusts cloud providers, their employees, and their software stack), but it is not zero trust.

\subsection{Isolation Model}

Hardware isolation in C8s sits at the CVM boundary, and where that boundary falls depends on the deployment. Where the boundary is drawn at the node, the CVM provides hardware-level isolation from the hypervisor and host OS, and pods inside a single CVM node are separated by standard Linux kernel mechanisms. Hard isolation is at the node boundary; pod-to-pod isolation on the same node is kernel-level, so a kernel exploit or misconfigured security context can bridge pods that share a node. Deployments requiring hardware isolation between tenants in this configuration should dedicate CVM nodes per tenant.

Where the boundary is drawn at the pod, the hardware boundary is the pod itself. Each pod's memory is encrypted with keys the hypervisor never possesses, and pods on the same physical host cannot reach each other's memory through a kernel exploit. Tenant isolation tightens to the pod level without requiring a node-per-tenant scheduling constraint.

\subsection{Operational Impact}

For platform operators already running Kubernetes, the primary operational changes are: (a) replacing standard node images with CVM node images, (b) deploying the CDS, NRI enforcer (DaemonSet), and raTLS mesh (DaemonSet) as infrastructure components, and (c) integrating the attested build pipeline (Kettle) into CI/CD. Existing deployments, Helm charts, and application code require no modification. The workload runs the same code in the same way; only the underlying node and network layer change.

C8s supports both all-CVM clusters and mixed clusters that combine CVM and non-CVM worker nodes under a single control plane. The per-pod raTLS mesh, NRI enforcer, and CDS operate unchanged on the CVM side; pods scheduled onto non-CVM nodes remain outside the trust boundary and do not receive mesh identity. Workload admission policy determines which workloads are permitted on which node type, letting operators run sensitive and non-sensitive workloads side by side in one cluster.

\subsection{Limitations}

The architecture has the following bounded scope:

\begin{itemize}
\item \textbf{GPU hardware without CC mode.} GPU-side protections are not provided on hardware that does not support confidential computing mode.
\item \textbf{Side channels.} Side-channel mitigations are outside the current scope.
\item \textbf{Runtime behavior.} The NRI image policy enforcer operates on image digests, not on runtime behavior; a vulnerability in an authorized image is exploitable within the TEE.
\item \textbf{Availability.} Availability guarantees are explicitly excluded; the architecture protects confidentiality and integrity, not uptime.
\end{itemize}

\section{Conclusion and Open Directions}
\label{sec:open}

The contribution of this paper is a single claim. Hardware-attested confidentiality for Kubernetes workloads is reachable without bringing the control plane inside the trust boundary, and the resulting architecture is compatible with the managed Kubernetes services (AKS, EKS, GKE) that most production deployments already run on. Section~\ref{sec:pod-cvm} onward describes the components that make this possible, including attestation-gated key brokering for sensitive artifacts such as model weights (\S\ref{sec:key-broker}).

A handful of directions sit on top of the architecture described here and are treated at depth elsewhere:

\begin{itemize}
\item \textbf{Attested build (Kettle).} A dedicated paper will cover the build environment, the attestation format, and the reproducibility guarantees.
\item \textbf{Finer-grained mesh termination when the boundary is at the pod.} The raTLS mesh currently terminates at the node in node-level deployments. When the boundary is drawn at the pod (the default; \S\ref{sec:pod-cvm}), the mesh can terminate inside the pod's CVM, tightening the boundary between workloads that share a host. The implementation and the incremental guarantee are under investigation.
\item \textbf{Per-workload measured configuration and per-customer signing rollout.} The measured-image-pinning and per-customer-signing mechanisms in \S\ref{sec:image-gap} are deployed for a small number of workloads today. Scaling signing to a platform-wide onboarding flow, caching measured configurations across workload classes, and sequencing key rotation without churning running pods are practical questions being worked through in production.
\item \textbf{Freshness of attestation.} TEE attestation throughput limits make per-connection attestation infeasible, which motivates the freshness beacon design in \S\ref{sec:default-tls}. Tightening the beacon window without saturating attestation throughput, and extending the same primitive to other verification points, remain open.
\item \textbf{Allow-list transparency log.} The CDS today exposes the current and previous allow-list from inside its attested CVM (\S\ref{sec:cds-deploy}). Extending this to a full append-only history with signed consistency roots, letting a third party verify that any past allow-list version is still in the log and that the log has not been rewritten, is a natural next step that would close the residual operator-side circularity.
\end{itemize}

Each of these sits on top of the architecture described here rather than changing it. The trust model, the threat model, and the design formula in \S\ref{sec:design-formula} are stable, and new work slots into them rather than revises them.

\section*{Acknowledgments}
We thank Ansgar Grunseid for contributions to design review and discussions, and Yolan Romailler for contributions to security review, proofreading, and design discussions.

\end{document}